\documentclass[preprint,floats,aps,12pt,superscriptaddress,nofootinbib,floatfix]{revtex4}
\usepackage{natbib} 
\usepackage{times} 
\usepackage{amssymb,amsmath}
\usepackage{csquotes}
\usepackage{prelim2e}\usepackage[none,bottom]{draftcopy}
\draftcopyName{preprint / }{1.2} 
\ifx\pdfoutput\undefined
\usepackage[usenames,dvips]{color}
\usepackage{graphicx}     
\else
\usepackage[usenames,dvipsnames]{color}
\usepackage{epstopdf}
\usepackage[pdftex]{graphicx}
\usepackage[pdftex]{hyperref}
\hypersetup{a4paper,
    pdftitle={Manuscript on XXX} 
    pdfauthor={et al. and Uwe Thiele},  
pdfproducer={lateX},
pdfview=FitV,       
pdfstartview=FitB,
linkcolor=blue,     
citecolor=blue,     
urlcolor=red,      
breaklinks=true,    
colorlinks=true,
citebordercolor=0 0 0,  
filebordercolor=0 0 0,
linkbordercolor=0 0 0,
menubordercolor=0 0 0,
urlbordercolor=0 0 0,
pdfhighlight=/I,
pdfborder=0 0 0,   
bookmarksopen=true,
bookmarksnumbered=true
}
\usepackage{braket}
\fi
\ifx\pdfoutput\undefined
\DeclareGraphicsExtensions{.eps}
\else
\DeclareGraphicsExtensions{.jpg, .pdf, .tif, .png}
\fi
\graphicspath{{.},{./}} 
\usepackage[usenames,dvipsnames]{color}
\usepackage[normalem]{ulem}

\newcommand{\Resub}[1]{{#1}}
\newcommand{\New}[1]{{#1}}
\newcommand{\Newest}[1]{{#1}}
\newcommand{\mylab}[1]{\label{#1}{\color{blue}\it \hspace*{.5cm} #1}}
\newcommand{\vecg}[1]{\boldsymbol{#1}}
\newcommand{\tens}[1]{\mathbf{\underline{#1}}}

\begin{document}
%
\title{Stationary broken parity states in \Resub{active matter} models}
\author{Tobias Frohoff-Hülsmann}
\email{t\_froh01@uni-muenster.de}
\thanks{ORCID ID: 0000-0002-5589-9397 }
\affiliation{Institut f\"ur Theoretische Physik, Westf\"alische Wilhelms-Universit\"at M\"unster, Wilhelm-Klemm-Str.\ 9, 48149 M\"unster, Germany}

\author{Max Philipp Holl}
\email{m.p.holl@uni-muenster.de}
\thanks{ORCID ID: 0000-0001-6451-9723 }
\affiliation{Institut f\"ur Theoretische Physik, Westf\"alische Wilhelms-Universit\"at M\"unster, Wilhelm-Klemm-Str.\ 9, 48149 M\"unster, Germany}
\author{Edgar Knobloch}
\email{knobloch@berkeley.edu}
\affiliation{Department of Physics, University of California at Berkeley, Berkeley CA 94720, USA}
\author{Svetlana V. Gurevich}
\email{gurevics@uni-muenster.de}
\thanks{ORCID ID: 0000-0002-5101-4686}
\affiliation{Institut f\"ur Theoretische Physik, Westf\"alische Wilhelms-Universit\"at M\"unster, Wilhelm-Klemm-Str.\ 9, 48149 M\"unster, Germany}
\affiliation{Center for Nonlinear Science (CeNoS), Westf{\"a}lische Wilhelms-Universit\"at M\"unster, Corrensstr.\ 2, 48149 M\"unster, Germany}
\author{Uwe Thiele}
\email{u.thiele@uni-muenster.de}
\homepage{http://www.uwethiele.de}
\thanks{ORCID ID: 0000-0001-7989-9271}
\affiliation{Institut f\"ur Theoretische Physik, Westf\"alische Wilhelms-Universit\"at M\"unster, Wilhelm-Klemm-Str.\ 9, 48149 M\"unster, Germany}
\affiliation{Center for Nonlinear Science (CeNoS), Westf{\"a}lische Wilhelms-Universit\"at M\"unster, Corrensstr.\ 2, 48149 M\"unster, Germany}
\affiliation{Center for Multiscale Theory and Computation (CMTC), Westf{\"a}lische Wilhelms-Universit\"at, Corrensstr.\ 40, 48149 M\"unster, Germany}
\begin{abstract}
We demonstrate that \Resub{several} nonvariational continuum models commonly used to describe active matter \New{as well as other active systems} exhibit nongeneric behavior: each model supports asymmetric but stationary localized states even \Resub{in the absence of pinning at heterogeneities. Moreover such states only begin to drift following a drift-transcritical bifurcation as the activity increases. Asymmetric stationary states should only exist in variational systems, i.e., in models with gradient structure. In other words, such states are expected in passive systems, but not in active systems where} the gradient structure of the model is broken by activity. We \New{identify a 'spurious' gradient dynamics structure of these models that is} responsible for this nongeneric behavior, and \Newest{determine} the types of additional terms that render the models generic, i.e., with asymmetric states that appear via drift-pitchfork bifurcations and are generically moving. We provide detailed illustrations of our results using numerical continuation of \New{resting and steadily drifting states in both generic and nongeneric cases.}
\end{abstract}
%
%
\maketitle
%
%
%
%
\section{Introduction}
The onset of spontaneous motion, i.e., the transition from a resting state to directed motion, is observed in many hydrodynamic~\cite{GGGC1991pra,SBL1988prl} systems. Examples include moving interfaces in directional solidification~\cite{SBL1988prl}, traveling spatially extended waves in a partially filled horizontal annulus with a rotating outer wall~\cite{MHAW1988pra}, traveling localized structures in binary fluid convection~\cite{NAC1990prl} and spatiotemporal chaos in directional viscous fingering \cite{RaMC1990prl}. Analogous behavior is also reported in e.g., optical~\cite{SCMG2019prl,JCMG2016prl,GMDS1990prl,AAS-PRE-96} and reaction-diffusion~\cite{KrischerPRL1994,OrGuilPRE98,P-PRL-01,Gurevich2004} systems. Furthermore, motion is essential for ``living'' systems like bacteria, animals, or (artificial) microswimmers which transform different forms of energy into self-propelled directed motion, giving rise to a wide variety of models for so-called active matter \cite{MJRL2013rmp,BFMR2022prx}. In these models the transition between resting and traveling states is of particular interest \cite{ZiSA2012jrsi,MeLo2013prl,SATB2014c,OpGT2018pre,StJT2022sm}.

The common property of all these systems is that they are permanently out-of-equilibrium, and so ``active''  as opposed to ``passive''. Here, passive systems are systems described by kinetic equations that have a variational structure \cite{Onsa1931pr1,Onsa1931pr2,Doi2011jpcm}, and can be written in the form of gradient dynamics on an underlying energy (or Lyapunov) functional $\mathcal{F}$. \New{In multicomponent systems the passive nature results in purely reciprocal interactions, i.e., couplings that obey the equivalent of Newton's third law for continuum models.}
Such models describe evolution towards stable or metastable equilibria corresponding to global and local minima of the energy functional, respectively. Thus no persistent time-periodic behavior such as traveling or standing oscillations is possible and all asymptotic states are time-independent, i.e., stationary. Important examples include the Allen-Cahn, Cahn-Hilliard, Swift-Hohenberg and phase field crystal models \cite{Cahn1965jcp,HoHa1977rmp,EKHG2002prl,BuKn2006pre,EGUW2019springer}.

In contrast, active systems are open systems as energy and/or matter flows through them: \New{for example, active particles may convert chemical energy into kinetic energy that} is continuously dissipated via friction \cite{MJRL2013rmp}. Such dissipative structures are \New{often described by continuum models corresponding to} kinetic equations that are nonvariational, i.e., at most part of the model can be brought into a gradient dynamics form and overall the form is broken. In consequence, time-periodic behavior and persistent motion become possible. 

In various models the nonvariational character is due to nonreciprocal coupling of the different order parameter fields \cite{MeLo2013prl,AgGo2019prl,SATB2014c,YoBM2020pna,SaAG2020prx,FrWT2021pre,FHLV2021n}, i.e., the various coupling terms cannot be consistently obtained as variations of a single energy functional \New{and the equivalent of Newton's third law for continuum models is broken~\cite{IBHD2015prx}.} Examples \New{where such models apply} include chemical or biophysical systems where \New{the effective chemical interaction between different species corresponds to a run-and-chase interaction scheme} while particles of the same species only interact passively, e.g.~via an attractive/repulsive interaction~\cite{AgGo2019prl}. In various recent studies the role of such nonreciprocal interactions in generating time-periodic behavior, suppressing coarsening and \New{inducing phase transitions between aligned and chiral phases} has been elucidated~\cite{FrWT2021pre,SaAG2020prx,YoBM2020pna,FHLV2021n}. 

Furthermore, many models of active systems assume a \Resub{linear nonvariational coupling between different order parameter fields} \cite{MeLo2013prl,SATB2014c,YoBM2020pna,SaAG2020prx,FrWT2021pre}. This assumption is related to the intuitive physical assumption that views active systems as perturbations of passive systems, resulting in a linear coupling at leading order. It is generally assumed that such linear, albeit nonvariational coupling suffices to capture the relevant qualitative behavior and that this behavior can then be related to the limiting passive system (or ``dead'' limit) whose thermodynamic behavior is often well understood~\cite{BFMR2022prx}. Here, we show that certain commonly used couplings in fact result in unexpected behavior in the onset of motion that indicates nongenericity of the resulting models.

In general, activity by itself does not automatically lead to motion. The onset of spontaneous motion is generally associated with broken parity symmetry, with the resulting asymmetry determining the direction and speed of propagation~\cite{CoGG1989prl,MHAW1988pra,StJT2022sm}. In translation-invariant systems a \New{primary} steady-state pattern-forming bifurcation creates a periodic state $u(\vec x)$ with definite parity under the spatial reflection $\vec x\to -\vec x $ \cite{CrKn1991arfm}: $u$ is even [odd] under this reflection, i.e., $u(\vec x)=u(-\vec x)$ [$u(\vec x)=-u(-\vec x)$]. 
In multicomponent systems, a mixed parity symmetry is also possible when certain components exhibit even parity while others have odd parity~\cite{OpGT2018pre}. Asymmetric states, i.e., states which \New{obey no parity symmetry},
are typically generated via a secondary, spontaneous parity-breaking bifurcation of a parity-symmetric state.\footnote{\New{Note that the state can have any parity symmetry, i.e., even, odd or mixed.}} In translation-invariant active systems such bifurcations generally lead to the onset of drift~\cite{GMDS1990prl,GoGG1990pra,GGGC1991pra,FaDT1991jp2} and are therefore referred to as drift-pitchfork bifurcations. \New{Ref.~\cite{KnTB1992pra} presents the corresponding normal form, i.e., the minimal set of ordinary differential equations that describes a (supercritical) drift-pitchfork bifurcation\footnote{\New{The normal form is  $\dot \theta = z$, $\dot z = \mu z-z^3$. The steady state $(z,\theta)=(0,\theta_0)$ represents the resting state that becomes unstable in the drift-pitchfork bifurcation at $\mu=0$ leading to the emergence of (left- and right) drifting states $(z,\theta)=(\pm \sqrt \mu,\pm \sqrt \mu t + \theta_0)$ for $\mu\geq0$. In fact, Ref.~\cite{KnTB1992pra} provides a more complete picture where a drift-pitchfork occurs as a secondary bifurcation. Then the minimal description is provided by $\dot r = \mu r - r^3$, $\dot \theta = z$, $\dot z = (r-1)z-z^3$ where the drift-pitchfork occurs at $\mu=1$, following a pitchfork bifurcation at $\mu=0$.}} and the emergence drifting structures in a two-component reaction-diffusion model. For partial differential equations of this type, Ref.~\cite{CoGG1989prl} proposes coupled amplitude and phase equations as universal lowest-order equations that capture the behavior in the vicinity of a drift-pitchfork bifurcation in translation- and reflection-symmetric systems.}
Since the drift-pitchfork bifurcation can be either supercritical or subcritical the transition to drift in experiments may occur in two different ways: in the supercritical case, the asymmetric state is stable at birth and its drift velocity increases continuously as the square root of the distance of the control parameter from its critical value. In contrast, in the subcritical case, the symmetric state changes dramatically on crossing the parity-breaking bifurcation, resulting in a state of finite asymmetry and speed. Furthermore, subcriticality creates bistability between asymmetric drifting and symmetric resting states, thereby permitting nucleation of inclusions of the broken-parity state within the symmetric one. \New{For spatially extended patterns} these spatiotemporal grain boundaries (defects) act as sources and sinks of traveling waves which locally propagate in opposite directions~\cite{GoGG1990pra,MiRa1992pd,OKGT2020c}, leading to wavelength relaxation~\cite{CoGG1989prl,GGGC1991pra}. 
\New{Drifting localized states can also emerge in drift-pitchfork bifurcations when the parity symmetry of a resting localized state is spontaneously broken. Such states are reported \Resub{in both one and higher spatial dimensions~\cite{Knob2015cmp,OKGT2021pre,AlCT2018c}}}.

\New{In the present work we are interested in {\it stationary} states of broken parity, i.e., resting asymmetric states. Besides variational systems, such states occur naturally in inhomogeneous systems due to pinning effects.  There, drift sets in via a depinning transition~\cite{Pome1986pd,TSTG2017pre} that is triggered when the degree of asymmetry of the stationary pinned state exceeds a threshold value. In contrast to homogeneous, i.e., translation-invariant systems, here the drift speed is no longer constant in time.}
Examples include stick-slip motion, e.g.~for droplets on an incline pinned by a chemical or topographic defect \cite{ThKn2006njp,BKHT2011pre}, or the orientation of liquid-crystal molecules when forced by a spatially periodic optical feedback \cite{HERB2010pre}. 

We consider resting asymmetric states in nonvariational reflection- and translation-invariant models. 
For these systems it is usually assumed that all broken parity states \textit{necessarily} drift~\cite{CoGG1989prl,Knob2015cmp}. In other words, all steady states in active systems have to obey a parity symmetry.
Note that this does not imply that such states cannot be time-periodic. Indeed, fixed-parity standing waves are also frequently found \cite{KnUY2021pre}. Although this assumption holds for scalar systems \cite{CrKn1991arfm}, recent work has identified several active systems described by the coupled dynamics of two or more order parameter fields that allow for \textit{parity-asymmetric steady states}. Moreover, the onset of drift may now occur via a drift-transcritical bifurcation.\footnote{\New{The normal form for this bifurcation is $\dot \theta = z$, $\dot z = \mu z-z^2$. The steady state $(z,\theta)=(0,\theta_0)$ represents the resting state that becomes unstable in the drift-transcritical bifurcation at $\mu=0$, giving rise to a branch of drifting states, $(z,\theta)=(\mu,\mu t + \theta_0)$, representing left-drifting states for $\mu < 0$ and right-drifting states for $\mu > 0$.}} As examples we mention steady localized asymmetric states in active phase field crystal (PFC) models \cite{OpGT2018pre,HAGK2021ijam,OKGT2021pre} and in Cahn-Hilliard (CH) models with nonreciprocal (i.e., nonvariational) coupling \cite{FrTh2021ima}. These states may or may not be stable, but as argued below, even stable states are not physically realizable owing to the limitation of these models as a realistic description of active systems.

The aim of our study is to understand why such states exist in particular models of active matter and reconcile this observation with the general expectation that all asymmetric states in nonvariational systems should move. We begin in Sec.~\ref{sec:generic} by reviewing the generic behavior and explaining how motion is related to asymmetry. In Sec.~\ref{sec:allmodels} we \New{introduce several examples of active models and for each case provide a bifurcation analysis that shows the existence of resting asymmetric states.} In particular, Secs.~\ref{sec:aPFCModel} and~\ref{sec:cCH} review and extend existing results for an active PFC model and a nonreciprocal two-field CH model, respectively, while Sec.~\ref{sec:CH+SH} discusses a model with \New{nonreciprocally} coupled conserved and nonconserved dynamics. Section~\ref{sec:a+pPFC} then provides results on a PFC model with three order parameter fields describing a mixture of passive and active particles while Sec.~\ref{sec:rd-system} considers a simple reaction-diffusion system. After presenting these individual examples, \New{Sec.~\ref{sec:existence} provides a general theory that identifies an entire class of active multicomponent models that allow for steady asymmetric states.} \New{We show that all these nonvariational models have a certain structure that allows bringing them into a form that resembles the gradient dynamics form of variational models, referred to as \textit{spurious gradient dynamics}. For this class of models we determine an explicit condition for the onset of motion.} 
In Sec.~\ref{sec:rewriteModels} we revisit \New{the models introduced in Sec.~\ref{sec:allmodels} and for each of them identify the ``hidden'' feature that is responsible for their spurious gradient structure and the resulting nongeneric behavior.} We also briefly discuss their relation to the skew-gradient form introduced in Refs.~\cite{Yana2002jde,KuYa2003pd}.
 We emphasize that despite their nongenericity the models we consider are quite standard and \New{appear in many previous studies through simplifications of more complicated continuum theories that are derived by coarse-graining of microscopic models. Alternatively they may be introduced phenomenologically using symmetry arguments or other purely macroscopic considerations.} \Newest{In either case the resulting models exhibit higher codimension dynamics not
found in the generic setting.} In Sec.~\ref{sec:restored} we employ a two-field coupled Swift-Hohenberg model to summarize the essential interplay \New{between parity symmetry and generic/nongeneric behavior} and use this example to explain how generic behavior can be restored. 
Finally, Sec.~\ref{sec:conc} discusses the implications of our work for the development of generic models for active systems. The data sets and plot functions for all figures as well as examples of \texttt{Matlab} codes for the employed numerical path continuation are provided on the open source platform \texttt{zenodo} \cite{FHKG2022zenodo}.

\section{Generic behavior}\label{sec:generic}

\New{We review the case of generic resting and drifting states in a one-component system in one dimension, with a specific parity symmetry representation. The argument below is generalized to multicomponent systems in more spatial dimensions and various parity symmetry representations in Appendix \ref{sec:general_parity}.}

 \New{We consider systems of the form
\begin{equation}
\partial_t u = F[u, \partial_x] \,,
\end{equation}
where $u(x,t)$ is an order parameter field that describes the evolution of the system in one spatial dimension $x$ and in time $t$. The function $F$ is taken to obey the parity symmetry $F[u,\partial_x]=F[u,-\partial_x]$. No further symmetries are assumed. Steady and stationary drifting states $u(x,t)=u^{(0)}(x-v t)$ with $v=0$ and $v\neq 0$, respectively, solve the steady state equation
\begin{equation}\label{eq:steady}
0=  F[u^{(0)}, \partial_\xi] + v \partial_\xi u^{(0)}\,,
\end{equation}
where $\xi=x- v t$ is the comoving frame variable.
Every profile $u^{(0)}(\xi)$ can now be written in the form
\begin{equation}\label{eq:uSuA}
u^{(0)}(\xi)=S u^{(0)}_\text{S}(\xi) + A u^{(0)}_\text{A}(\xi)
\end{equation}
with a normalized symmetric part $u^{(0)}_\text{S}(-\xi)=u^{(0)}_\text{S}(\xi)$ and a normalized antisymmetric part $u^{(0)}_\text{A}(-\xi)=-u^{(0)}_\text{A}(\xi)$ and their respective amplitudes $S$ and $A$. We say that $u(\xi)$ has even [odd] parity or is symmetric [antisymmetric] under reflection if $A=0$ [$S=0$].\footnote{Note that for the existence of antisymmetric steady states $F$ has to obey a further inversion symmetry, namely, be antisymmetric with respect to the order parameter $u$, i.e., $F[-u,\partial_x]=-F[u,\partial_x]$.} Here, we consider a symmetric state $u^{(0)}(\xi)=S u^{(0)}_\text{S}(\xi)$, i.e., $A=0$, and apply the parity transformation $\xi\to -\xi$ to the steady state equation~\eqref{eq:steady}. This yields
\begin{align}
0=&  F[S u^{(0)}_\text{S}(-\xi), -\partial_\xi] - v \partial_\xi S u^{(0)}_\text{S}(-\xi)= F[S u^{(0)}_\text{S}(\xi), \partial_\xi] - v \partial_\xi S u^{(0)}_\text{S}(\xi)\,,
\end{align}
where in the final step we used the parity symmetry of $u^{(0)}_\text{S}$ and of $F$.
Comparing with Eq.~\eqref{eq:steady} we conclude that the velocity $v$ vanishes. Thus, parity-symmetric steady states do not drift.\footnote{Antisymmetric states $u^{(0)}_\text{A}$ are likewise necessarily at rest if they exist, i.e., if the inversion antisymmetry $F[u^{(0)}, \partial_\xi]=-F[-u^{(0)}, \partial_\xi]$ holds.} 
}

\New{Next, we consider an asymmetric state where the antisymmetric contribution has a small amplitude $A\ll 1$ that may be controlled by an appropriate parameter.  Expanding the steady state Eq.~\eqref{eq:steady} in $A$ yields
\begin{equation}\label{eq:symm1}
0= F[S u^{(0)}_\text{S}(\xi), \partial_\xi]  + A  J[S u^{(0)}_\text{S}(\xi), \partial_\xi]  u^{(0)}_\text{A}(\xi) + v \partial_\xi \left(S u^{(0)}_\text{S}(\xi) + A u^{(0)}_\text{A}(\xi)\right) + \mathcal{O}(A^2)\,.
\end{equation}
where $J[S u^{(0)}_\text{S}(\xi), \partial_\xi]$ is the differential Jacobi operator, i.e., the one-dimensional Jacobi matrix in spatial representation.
Applying the parity transformation $\xi\to -\xi$ gives
\begin{align}
  0= F[S u^{(0)}_\text{S}(-\xi), -\partial_\xi]  + A\,J[u^{(0)}_\text{S}(-\xi), -\partial_\xi]  u^{(0)}_\text{A}(-\xi) - v \partial_\xi \left(S u^{(0)}_\text{S}(-\xi) + A u^{(0)}_\text{A}(-\xi)\right) + \mathcal{O}(A^2).\nonumber
\end{align}
Using the symmetries of $u^{(0)}_\text{S}$, $u^{(0)}_\text{A}$ and $F$ we obtain
\begin{align}
 \Rightarrow 0 =  F[S u^{(0)}_\text{S}(\xi), \partial_\xi]  - A\,J[u^{(0)}_\text{S}(\xi), \partial_\xi]  u^{(0)}_\text{A}(\xi) - v \partial_\xi \left(S u^{(0)}_\text{S}(\xi) - A u^{(0)}_\text{A}(\xi) \right)+ \mathcal{O}(A^2)\,, \label{eq:symm2}
\end{align}
where we used that $J[u, \partial_\xi]$ inherits the parity symmetry of $F[u, \partial_x]$, i.e., $J[u, -\partial_\xi]=J[u, \partial_\xi]$.
Comparison of Eqs.~\eqref{eq:symm1} and~\eqref{eq:symm2} gives
\begin{equation}\label{eq:symm3}
0= A\,J[u^{(0)}_\text{S}(\xi), \partial_\xi]  u^{(0)}_\text{A}(\xi) + S v \partial_\xi u^{(0)}_\text{S}(\xi) + \mathcal{O}(A^2)\,.
\end{equation}
Multiplying Eq.~\eqref{eq:symm3} by $\partial_\xi u^{(0)}_\text{S}(\xi)$ and integrating over the domain yields the propagation velocity $v$ at leading order,
\begin{equation}\label{eq:v}
v= - \frac{A}{S}\, \frac{\int_{-L/2}^{L/2} {\rm d} \xi \, \partial_\xi u^{(0)}_\text{S}(\xi) J[u^{(0)}_\text{S}(\xi), \partial_\xi]  u^{(0)}_\text{A}(\xi)}{\int_{-L/2}^{L/2} {\rm d} \xi \left(\partial_\xi u^{(0)}_\text{S}(\xi)\right)^2}\,.
\end{equation}
For steady states with a large antisymmetric part and a small symmetric part (i.e., $S\ll1$) an equivalent expression holds, with only the amplitudes $A$ and $S$ and the subscripts interchanged.
Moreover, as elaborated in Appendix~\ref{sec:general_parity}, for a general parity symmetry in a $d$ dimensional system described by a multicomponent order parameter field $\vecg u$, the velocity components $v_i$ for $i=1,\dots,d$ are given by 
\begin{equation}\label{eq:v_general}
v_i=- \mu\, \frac{\left< \partial_{\xi_i} \vecg{u^{(0)}} |\tens J \vecg{u^{(1)}}\right>}{\left<\partial_{\xi_i} \vecg{u^{(0)}}| \partial_{\xi_i} \vecg{u^{(0)}}\right>}=- \mu\,\frac{\left< \tens J^\dagger \partial_{\xi_i} \vecg{u^{(0)}} | \vecg{u^{(1)}}\right>}{\left<\partial_{\xi_i} \vecg{u^{(0)}}| \partial_{\xi_i} \vecg{u^{(0)}}\right>}\,,
\end{equation}
where $\mu$ denotes the ratio of the amplitude of the asymmetric contribution $\vecg{u^{(1)}}$ to the amplitude of the parity-symmetric part $\vecg{u^{(0)}}$ of the steady state $\vecg{u^{(0)}}$. This equation generalizes the expression in Eq.~\eqref{eq:v} using $\left<\dots\right>$ to denote a scalar product, i.e., a product in the space of order parameter fields followed by an integration over the domain; $\partial_{\xi_i} $ denotes the spatial derivative in the $\xi_i$ direction. 
The numerator of Eq.~\eqref{eq:v_general} is sometimes called the degree of asymmetry~\cite{CoGG1989prl}.
Thus, in general, asymmetric states drift with a speed that is proportional to the degree of asymmetry and in particular proportional to the amplitude ratio $\mu$ that may serve as the control parameter in the normal form of the drift-pitchfork bifurcation.}

\New{For variational, nonconserved dynamics the Jacobi matrix is self-adjoint, i.e., $\tens J^\dagger = \tens J$. Moreover $\partial_{\xi_i} \vecg{u^{(0)}}$ is the Goldstone mode that is related to the translation symmetry of the model. Thus
\begin{equation}
\tens J\,\,\partial_{\xi_i} \vecg{u^{(0)}}=0 \, \, \forall i
\end{equation}
and the numerator in Eq.~\eqref{eq:v_general} vanishes, a result that is consistent with the absence of long-time dynamics in such systems.
In contrast, in nonvariational, i.e., active systems, the velocity $v$ given by Eq.~\eqref{eq:v_general} is no longer identically zero and motion is therefore expected, except possibly at isolated parameter values. In such systems branches of asymmetric resting states are therefore nongeneric despite their presence in various commonly used active models. This apparent contradiction represents the starting point for this work and is resolved here by identifying the special structure of these models, inadvertently introduced at various points in their construction. Our considerations indicate, furthermore, that in the generic case drift sets in via drift-pitchfork bifurcations corresponding to spontaneous parity breaking while in nongeneric systems the onset of drift is not linked to spontaneous symmetry breaking, and so arises via drift-transcritical bifurcations as described in~\cite{OpGT2018pre} and further elaborated in the following sections.}

\section{Nongeneric behavior: Asymmetric steady states in active models} \label{sec:allmodels}
In this section we present several examples of nonvariational models where in contrast to expectations (Sec.~\ref{sec:generic}) resting parity-asymmetric states are found. We have selected representatives of different classes of models, focusing on models with conserved, nonconserved and mixed dynamics. Our first two examples, an active PFC model in Sec.~\ref{sec:aPFCModel} and an active two-field Cahn-Hilliard model with nonreciprocal interactions in Sec.~\ref{sec:cCH}, exhibit conserved dynamics for the density fields (and mixed dynamics for the polarization field in the PFC case). Next, in Sec.~\ref{sec:CH+SH} we consider a model for the interaction between a large-scale (i.e., long-wave) and small-scale (i.e., finite wavelength) instability represented by nonreciprocally coupled Cahn-Hilliard and Swift-Hohenberg equations. The behavior of the two-variable models in Secs.~\ref{sec:aPFCModel}--\ref{sec:CH+SH} is then compared in Sec.~\ref{sec:a+pPFC} to a three-variable PFC system consisting of two coupled conserved density fields, \New{one of which represents active particles and therefore couples to a polarization field (rendering the model active).}
All these models commonly arise in the description of pattern formation in physico-chemical and biophysical nonequilibrium systems. 
We conclude the presentation of individual models in Sec.~\ref{sec:rd-system} by showing that such nongeneric behavior also appears in a standard reaction-diffusion model with nonconserved dynamics, namely, the FitzHugh--Nagumo model.

All the bifurcation diagrams shown below are obtained using numerical continuation \cite{EGUW2019springer} employing the \texttt{Matlab} package \texttt{pde2path} \cite{UeWR2014nmma}. As a solution measure we always use the L$_2$ norm
\begin{equation}\label{eq:norm}
||\delta u|| = \sqrt{ \sum_i  \frac{1}{L}\int_{-\frac{L}{2}}^\frac{L}{2} {\rm d}x \left(u_i- \bar u_i\right)^2}\,,
\end{equation}
where the $u_i$ correspond to the order parameter fields of the model, the $\bar u_i$ are their mean values and $L$ is the domain size.
\New{Furthermore, we consistently use markers and line styles to indicate different bifurcations and states as listed in Table~\ref{tab:instab}. In particular, drift bifurcations are indicated by triangle symbols, empty ones for generic drift-pitchfork bifurcations and filled ones for nongeneric drift-transcritical bifurcations. The resulting branches of stationary drifting states are indicated by brown dashed dotted lines if unstable. Stable states, whether steady (resting) or not, are always indicated by solid lines. We use line colors to distinguish between different branches of steady states. 
\begin{table}[hbt]
\begin{tabular}{|c | c || c | c|}
\hline
Bifurcation & Symbol & Branch & Line style\\
\hline
\hline
pitchfork bifurcation
&
\includegraphics{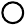}
& 
stable states
&
\includegraphics[height=0.02\textheight]{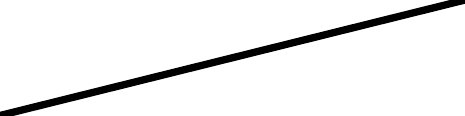}
~\\
\hline
drift-pitchfork bifurcation
&
\includegraphics{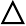}
& 
unstable steady states
&
\includegraphics[height=0.02\textheight]{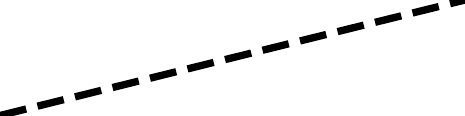}
~\\
\hline
drift-transcritical bifurcation
&
\includegraphics{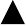}
& 
unstable stationary drifting states
&
\includegraphics[height=0.02\textheight]{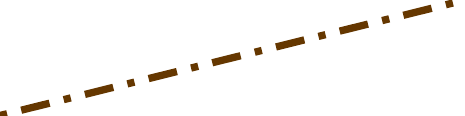}
~\\
\hline
Hopf bifurcation
&
\includegraphics{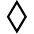}
& 
unstable standing waves
&
\includegraphics[height=0.02\textheight]{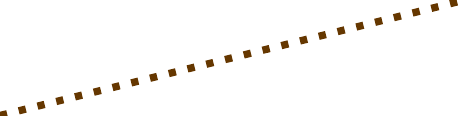}
~\\
\hline
\end{tabular}
\caption{List of symbols and line styles used in all bifurcation diagrams to differentiate between different types of bifurcations and solution branches.}
\label{tab:instab}
\end{table}}

\subsection{Active phase field crystal model}\label{sec:aPFCModel}

Our first model is the active PFC model studied in Refs.~\cite{MeLo2013prl,MeOL2014pre,ChGT2016el,OpGT2018pre,OKGT2020c,OKGT2021pre}. In the passive limit, the PFC model (aka conserved Swift-Hohenberg model) provides a simple mean-field description of crystallization processes \cite{EKHG2002prl,ELWG2012ap,TARG2013pre} in terms of the gradient dynamics of a single conserved order parameter field, the density $\phi(\vec{x},t)$. The underlying free energy $\mathcal{F^{\text{SH}}}$ is of Swift-Hohenberg type \cite{CrHo1993rmp, BuKn2006pre}. This passive model is made active by coupling the density dynamics to the dynamics of a (vector) polarization field $\vec{P}(\vec{x},t)$~\cite{MeLo2013prl}. The linear coupling employed represents the leading order active term and can be derived, \Resub{like the entire model}, from a microscopic dynamical density functional theory (DDFT) via a combined gradient and Taylor expansion~\cite{MeOL2014pre,VrLW2020aip}. \Resub{Note that DDFT itself corresponds to a microscopic continuum description as it can be systematically obtained from the equations of motion of individual particles \cite{VrLW2020aip}.}

In one spatial dimension the active PFC model reads
\begin{alignat}{1}
\begin{aligned}
\partial_{t}\phi  &=  \partial_{xx} \frac{\delta \mathcal{F^{\text{SH}}}}{\delta \phi} -v_{0}\partial_x P\,, ~\\
\partial_{t} P &= \partial_{xx}\frac{\delta \mathcal{F}^P}{\delta P} - D_{\mathrm{r}}\frac{\delta \mathcal{F}^P}{\delta P} - v_{0} \partial_x \phi
\end{aligned}
\label{eq:apfc_functional}
\end{alignat}
with the functionals
\begin{alignat}{1}
\begin{aligned}
\mathcal{F^{\text{SH}}} =& \int_{-\frac{L}{2}}^{\frac{L}{2}} {\rm d} x  \biggl[\frac{\phi}{2}\left[\epsilon+\left(1+\partial_{xx}\right)^{2}\right]\phi+ \frac14 \phi^{4}\biggr]\,,~\\
\mathcal{F}^P =& \int_{-\frac{L}{2}}^{\frac{L}{2}} {\rm d} x  \biggl[\frac{1}{2} c_1 P^2 + \frac{1}{4} c_2 P^4\biggr]\,.
\end{aligned}
\label{eq:apfc_freeenergies}
\end{alignat}
Here the coupling parameter $v_{0}$ represents activity and is referred to as the self-propulsion velocity. The dynamics of the polarization $P(x,t)$ contains translational and rotational diffusion captured by mixed conserved and nonconserved gradient dynamics, with $D_\mathrm r$ being the rotational diffusivity. \New{The free energy of the density field $\phi$ depends on the effective temperature parameter $\epsilon$ and defines a characteristic length scale, set to $1$. The mean density $\bar \phi\equiv (1/L) \int_{-L/2}^{L/2}{\rm d}x\,\phi$ can also be used as a parameter.\footnote{\New{This is made clear on using the order parameter $\tilde \phi\equiv\phi - \bar \phi$ instead of $\phi$. In this formulation the mean density of $\tilde \phi$ is zero, but $\bar \phi$  explicitly appears as a parameter in the dynamic equation.}}} The free energy density of the polarization field represents a parabolic (for $c_1>0$) or a double-well potential (for $c_1<0$ and $c_2 \geq 0$). In the former case diffusion reduces polar order, in the latter case spontaneous polarization arises. Computing the functional derivatives in Eqs.~\eqref{eq:apfc_functional} yields
\begin{alignat}{1}
  \begin{aligned}
  \partial_{t}\phi &= \partial_{xx}\left\{\left[\epsilon+\left(1+\partial_{xx}\right)^{2}\right]\phi+\phi^{3}\right\}-v_{0}\partial_x P\,, ~\\
\partial_{t} P &= \partial_{xx}\left(c_1 P + c_2 P^3\right) - D_{\mathrm{r}}(c_1 P +  c_2 P^3) - v_{0} \partial_x \phi\,.
\end{aligned}
\label{eq:dtapfc}
\end{alignat}
Based on the scalar and vector character of $\phi$ and $P$, respectively, the model~\eqref{eq:dtapfc} is invariant under the parity symmetry transformation $\mathcal{R}: \left(x, \phi, P\right) \mapsto \left(-x, \phi, -P\right)$. According to this symmetry, steady states that are invariant under $\mathcal{R}$, i.e., with $\left(\phi(x), P(x)\right) = \left(\phi(-x), - P(-x)\right)$, are referred to as parity-symmetric, with \New{mixed parity.} Note that if $\bar \phi=0$ the model also obeys inversion symmetry $\mathcal{I}: (\phi,P) \mapsto (-\phi,- P)$.

\begin{figure}
	\includegraphics[width=\textwidth]{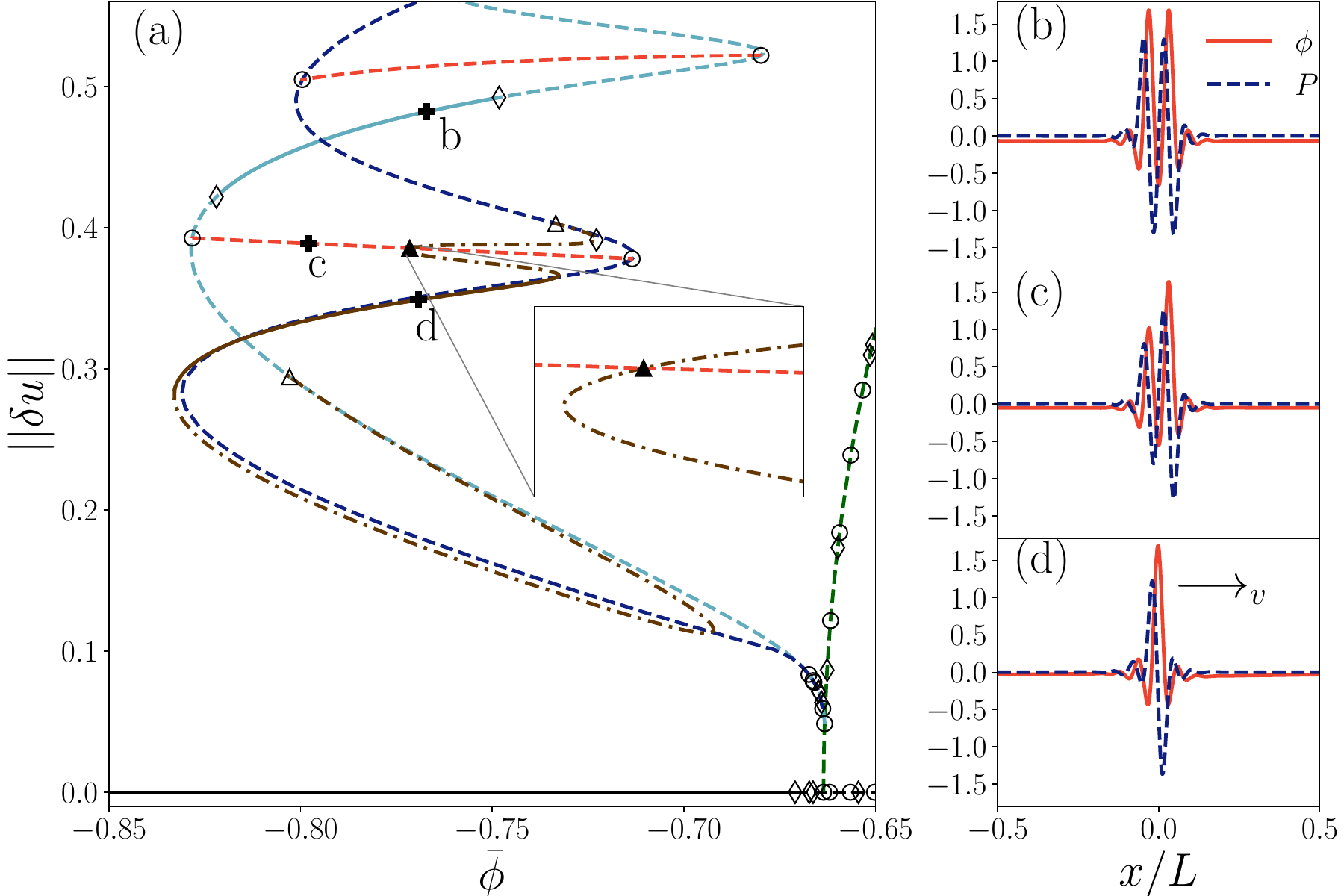}
	\includegraphics[width=\textwidth]{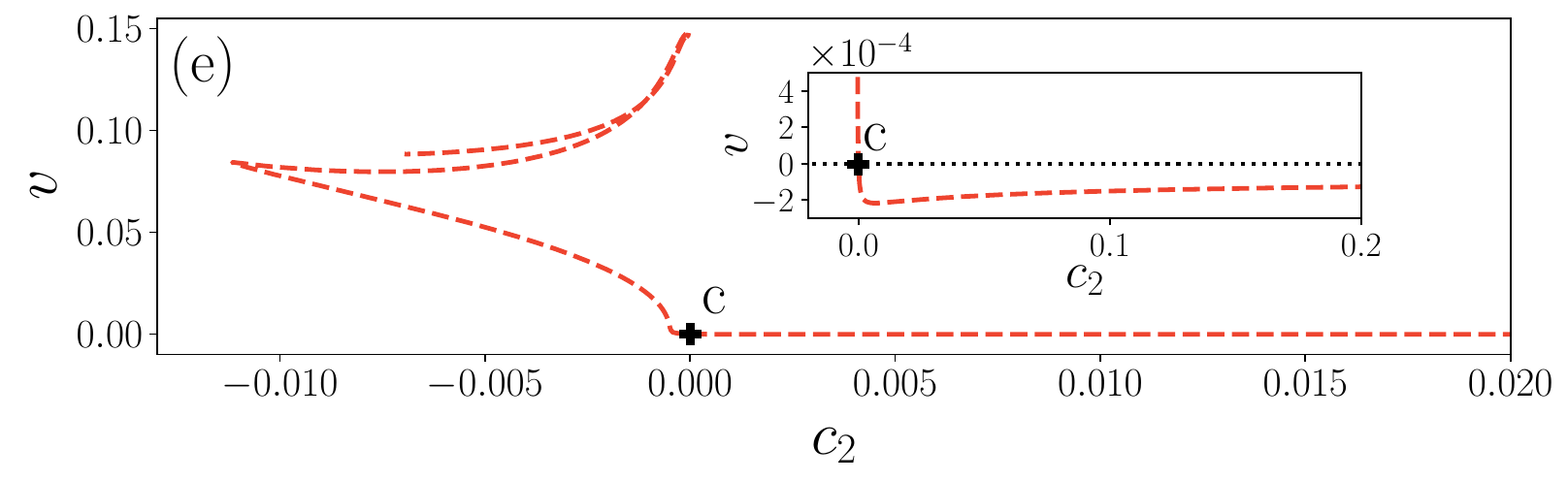}
	\caption{Localized states in the active phase field model \eqref{eq:dtapfc}. (a) Bifurcation diagram showing the L$_2$ norm~\eqref{eq:norm}
with $u_1=\phi$ and $u_2=P$ as a function of the parameter $\bar \phi$. The primary bifurcation from the trivial state (black line) generates periodic states (dark green line), followed by a secondary bifurcation to a pair of stationary parity-symmetric localized states with odd and even numbers of peaks (resp., RLS$_\text{odd}$ and RLS$_\text{even}$) are shown as dark and light blue lines. Tertiary branches of asymmetric resting rung states (red lines) and asymmetric traveling states (brown line) are also shown. 
\New{Line styles and symbols are employed as listed in Table~\ref{tab:instab}.}
Panels (b)-(d) show sample profiles at the locations indicated by crosses in panel~(a). Parameters: $\varepsilon=-1.5, c_1=0.1, D_r=0.5$, $c_2=0$, $v_0 = -0.16475$ (note that the sign of $v_0$ is given incorrectly in \cite{OpGT2018pre}). The domain size is $L=100$. Panel~(e) shows that the steady rung state of panel~(c) starts to move when $c_2\neq 0$.
}
\label{fig:aPFC}
\end{figure}

For $v_0=0$, the two equations~\eqref{eq:dtapfc} decouple and one recovers the (passive) PFC model. In contrast, for $v_0\neq 0$ the system is active and can no longer be written in the form of gradient dynamics. Thus for $v_0\neq 0$ we expect to find time-dependent dynamics and, in particular, traveling structures, resulting in a rich variety of complex behavior \cite{MeLo2013prl} that can be represented in intricate bifurcation diagrams \cite{OpGT2018pre,OKGT2020c}. Here, we focus on a subset of the extensive results presented in \cite{OpGT2018pre}, namely, on the bifurcation behavior of the resting localized states 
 as summarized in Fig.~\ref{fig:aPFC}. 

Figure~\ref{fig:aPFC}(a) shows branches of resting and traveling localized states that form a slanted snaking bifurcation structure~\cite{Knob2016ijam,TARG2013pre} where solid and dashed lines indicate linearly stable and unstable states, respectively. The snaking structure consists of two intertwined branches of steady parity-symmetric states RLS$_\text{odd}$ and RLS$_\text{even}$ with an odd (dark blue) or even [light blue, see e.g.~profile in Fig.~\ref{fig:aPFC}(b)] number of peaks. \New{As is typical for mass-conserving systems the snaking structure is slanted or tilted, in contrast to non-mass-conserving systems where the structure is vertical~\cite{BuKn2007pla}.}
The interconnecting rung-like branches (red) consist of steady asymmetric states [see e.g.~profile in Fig.~\ref{fig:aPFC}(c)]. \New{For the passive PFC model ($v_0=0$), the gradient structure of the model allows for the presence of the steady asymmetric rung states as explained in Sec.~\ref{sec:generic}. However, it is surprising that these states remain at rest also in the active case $v_0\neq 0$.} Beside steady states one also finds branches of traveling states [brown branches, see e.g.~profile in Fig.~\ref{fig:aPFC}(d)] that bifurcate from the branches of steady symmetric states via standard drift-pitchfork bifurcations (empty triangle symbol), and from the steady asymmetric states via the nonstandard drift-transcritical bifurcation [filled triangle symbol, see inset in Fig.~\ref{fig:aPFC}(a) for a magnification]. 
\New{In the former case two branches of symmetry-related traveling states (left- and right-traveling) emerge from the branch of resting states, thereby breaking the (mixed) parity symmetry of the resting states. As a result the two emerging branches correspond to states that are related by the parity symmetry $\mathcal{R}$. In the latter case, the resting states are already asymmetric and so also lie on a pair of distinct branches with states of opposite asymmetry. However, this time the states lose stability to drifting asymmetric states at drift-transcritical bifurcations and the drift direction and speed depend on the asymmetry of the steady state generating these states.}

We emphasize that this behavior is surprising: contrary to the expectation that, \Resub{generically,} asymmetric states drift in all active systems, here branches of resting and traveling asymmetric states coexist and, moreover, are related via drift-transcritical bifurcations that should not exist in generic systems. \New{These facts suggest that the model \eqref{eq:dtapfc} is nongeneric.}

Figure~\ref{fig:aPFC}(a) further reveals the presence of several Hopf bifurcations, some on the branches of even and odd localized states, and one more on the branch of drifting asymmetric states (open diamond symbols). The latter generates a quasiperiodic state. The Hopf bifurcations on RLS$_\text{odd}$ and RLS$_\text{even}$ are created via Bogdanov-Takens bifurcations and so first arise with zero frequency at the folds of the slanted snaking structure. With increasing activity $v_0$ these bifurcations move inwards, decreasing the parameter range where RLS are linearly stable, and may ultimately annihilate, thereby rendering all steady localized states unstable. We have not explored the resulting dynamical states, but refer to Sec.~\ref{sec:cCH} for a detailed study of a similar problem.

\New{Next, we take a resting asymmetric state in Fig.~\ref{fig:aPFC}(a) and increase the strength $c_2$ of the higher order polarization term from zero. Usually, $c_2>0$ is demanded to avoid blow-up in the polarization, particularly for $c_1<0$. Here, we use $c_1>0$ which allows us to also investigate the behavior for small negative values of $c_2$.} Panel~(e) shows the velocity $v$ of the asymmetric state in panel~(c) as a function of $c_2$ and shows that the state begins to move as soon as $c_2 \neq 0$. For $c_2>0$ the velocity $v$ first decreases, then increases again, but remains small and negative while approaching a plateau (see the inset). For $c_2<0$ the velocity is positive and much larger; $v$ increases monotonically with decreasing $c_2$ until a fold where two branches of traveling states merge. Following the upper branch for increasing $c_2$ we find that the branch approaches $c_2=0$ where $v\approx 0.15$ before doubling back. The resulting complex behavior is omitted from the plot. \New{The key point is that the aforementioned nongeneric behavior appears to be lifted for any $c_2\neq0$, a fact that demands explanation (see Sec.~\ref{sec:apfc} below). In the remainder of this section we pursue the question whether the nongeneric behavior identified above is exclusive to the active PFC model or is a common feature also of other active models.}

\subsection{Nonreciprocal Cahn-Hilliard model}\label{sec:cCH}

The nonreciprocal Cahn-Hilliard model has been analyzed in several recent studies~\cite{SaAG2020prx,YoBM2020pna,FrWT2021pre,FrTh2021ima,BFMR2022prx} and provides a description of mixtures of nonreciprocally interacting (i.e., active) colloids. The model is based on the Cahn-Hilliard equation, a passive model originally proposed to describe phase separation of isotropic solid or binary fluid phases~\cite{CaHi1958jcp,Cahn1965jcp}. Owing to mass conservation the model takes the form of a pair of coupled continuity equations driven by gradients in the corresponding chemical potentials. Since it captures many qualitative features of phase separation the model and its variants and extensions are widely applied in, e.g., biophysics and soft matter contexts. Variants can be distinguished by the features of the original model that are modified or broken. For example, in the convective Cahn-Hilliard model the parity symmetry is broken by a directed driving force accompanied by a flux across the system boundaries \cite{WORD2003pd,TALT2020n}. Nonvariational extensions of the model that preserve parity symmetry are used to describe motility-induced phase separation of active Brownian particles \cite{SBML2014prl,RaBZ2019epje}. 

The nonreciprocal Cahn-Hilliard model describes the dynamics of two order parameter fields $\phi_1$ and $\phi_2$ that represent scaled and shifted densities. Cahn-Hilliard equations for the individual fields are augmented by a linear coupling resulting in
	\begin{alignat}{1}
	\begin{aligned}
	\qquad \frac{\partial\phi_1}{\partial t}  &= \frac{\partial^2}{\partial x^2} \left(- \frac{\partial^2\phi_1 }{\partial x^2} +f_1'(\phi_1) - \left(\rho + \alpha\right) \phi_2 \right)\,,\\
	\qquad \frac{\partial\phi_2}{\partial t}  &=  \frac{\partial^2}{\partial x^2} \left(-\kappa\frac{\partial^2 \phi_2}{\partial x^2} +f_2'(\phi_2) - \left(\rho - \alpha\right) \phi_1 \right)\,,
	\end{aligned}
	\label{eq:CHCH}
	\end{alignat}
where
	\begin{alignat}{1}
	\begin{aligned}
	f_1=&  \frac{1}{2}a \phi_1^2  + \frac{1}{4} \phi_1^4\,,\\
	f_2=& \frac{1}{2}(a+a_\Delta) \phi_2^2  + \frac{1}{4} \phi_2^4\,
	\end{aligned}
	\label{eq:CHCH_ff}
	\end{alignat}
represent double-well potentials. Mass conservation implies that at all times
	\begin{alignat}{1}
	\begin{aligned}
	\frac{1}{L}\int_{-\frac{L}{2}}^{\frac{L}{2}} {\rm d} x \,\phi_1 = & \bar \phi\,,\\
	\frac{1}{L}\int_{-\frac{L}{2}}^{\frac{L}{2}} {\rm d} x \,\phi_2 = & \bar \phi + \Delta \bar \phi\,,
	\end{aligned}
	\label{eq:CHCH_int}
	\end{alignat}
where $L$ is the fixed domain size and $\bar \phi$ and $\bar \phi + \Delta \bar \phi$ represent constant mean densities.
        
The model \eqref{eq:CHCH} is invariant with respect to $\mathcal{R}: (x,\phi_1,\phi_2) \mapsto (-x,\phi_1,\phi_2)$ and, if both mean densities vanish, i.e., if $\bar \phi=\Delta \bar \phi=0$, it is also inversion-symmetric, i.e., invariant under $\mathcal{I}: (\phi_1,\phi_2)\mapsto (-\phi_1,-\phi_2)$. Compared to the active PFC model of the previous section the parity symmetry $\mathcal{R}$ takes a different representation due to the scalar character of both density fields in the model. Hence, parity-symmetric states are those that are invariant under $\mathcal{R}$, while in the special case $\bar \phi=\Delta \bar \phi=0$ we also have parity-antisymmetric states, i.e., states invariant under $\mathcal{R} \circ \mathcal{I}$.

\begin{figure}[h!]
		\includegraphics[width=\textwidth]{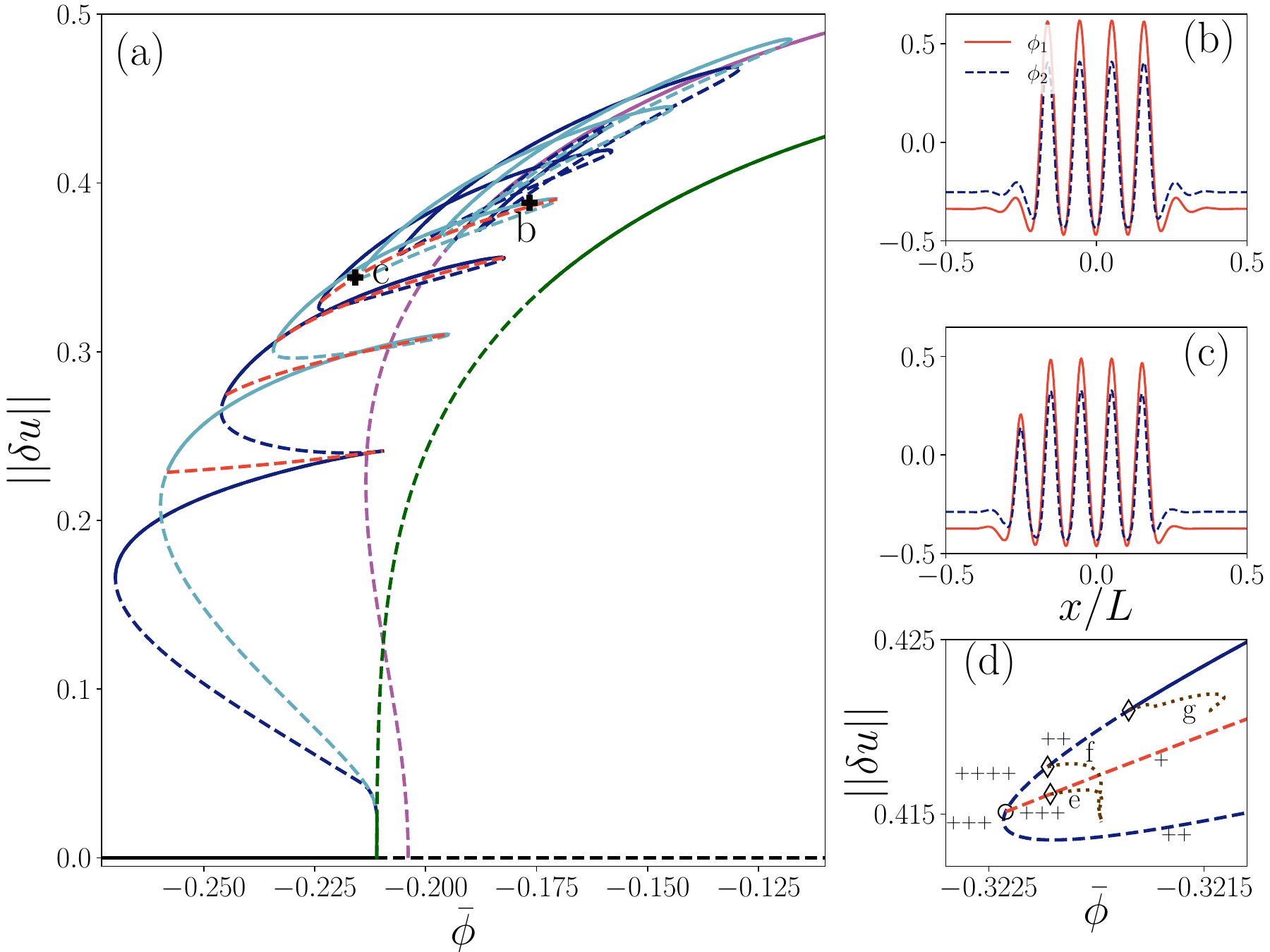}
		\includegraphics[width=\textwidth]{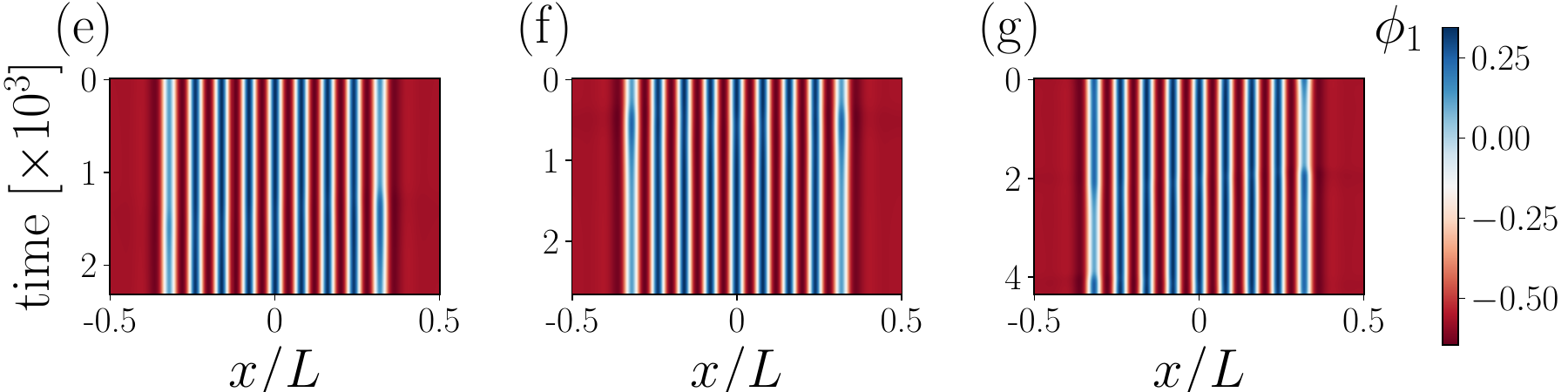}
		\caption{(a) The L$_2$ norm for the system \eqref{eq:CHCH} as a function of the mass $\bar \phi$ for $\alpha = 1.65$, $\Delta\bar \phi=0$. Steady symmetric localized states with odd and even numbers of peaks (resp., RLS$_\text{odd}$ and RLS$_\text{even}$) are shown as dark and light blue lines. Also included are the four lowest branches of steady asymmetric localized states (red dashed lines), the periodic steady states with $n=11$ peaks (dark green line) and $n=9$ peaks (light purple line), and the uniform state (black line). Panels (b) and (c) show sample profiles at parameters marked by bold symbols in (a). The remaining parameters are $\kappa=0.14$, $a=1.25$, $a_\Delta = -1.9$, $\rho = 1.35$. The domain size is $L=20 \pi$. Panel~(d) shows the existence of Hopf bifurcations (diamond symbols) near the ninth fold of the RLS$_\text{odd}$ branch for $\alpha=1.78$ and $a=0.9$ and unchanged values of the other parameters. The resulting branches of standing oscillations are represented by dotted brown lines, with sample solutions shown in panels~(e)-(g). The number of unstable eigenvalues along the steady state branches is indicated by $+$ symbols. All line styles and symbols are as in Table~\ref{tab:instab}.}
		\label{fig:CH+CH_snaking} 
	\end{figure}

The model \eqref{eq:CHCH} is referred to as ``nonreciprocal'' since the parameter $\alpha$, the antisymmetric part of the coupling, represents a ``run-and-chase'' interaction between the two fields/species $\phi_1$ and $\phi_2$, i.e., it breaks Newton's third law \cite{IBHD2015prx}. Here, we revisit recently published results on localized structures in this model~\cite{FrTh2021ima} and focus on the presence of asymmetric steady states in the active case $|\alpha|>|\rho|$. Figure \ref{fig:CH+CH_snaking}(a) presents a bifurcation diagram for $\Delta\bar \phi=0$ using the mean density $\bar \phi$ as a control parameter. Solid and dashed lines indicate linearly stable and unstable states, respectively; dotted lines indicate unstable time-periodic states. The reciprocal and nonreciprocal interaction strengths are $\rho = 1.35$ and $\alpha=1.65$, respectively. In this case, a linear stability analysis reveals that the homogeneous state exhibits a small-scale instability~\cite{FrTh2021ima} leading to a supercritical branch of stationary periodic states (dark green). Owing to its spatial period we refer to this branch as the $n=11$ branch. However, this branch loses stability almost immediately at a secondary bifurcation generating two branches of stationary symmetric localized states RLS$_\text{odd}$ (dark blue) and RLS$_\text{even}$ (light blue) where as in the previous section the subscripts refer to the number of peaks. These localized states emerge subcritically and are organized in a slanted snaking structure with alternating stable segments, much as found for the active PFC model (Fig.~\ref{fig:aPFC}); every second fold one pattern wavelength is added symmetrically on either side of the structure, resulting in the addition of two new peaks. This process continues as one follows RLS$_\text{odd}$ and RLS$_\text{even}$ to larger norm until the whole domain is filled with peaks [see e.g.~the profile of the four peak state in panel (b)] and the RLS branches terminate on a branch of periodic states, here the $n=9$ branch (light purple). Within the slanted snaking structure one finds interconnecting branches of asymmetric rung states that emerge and terminate in pitchfork bifurcations near the RLS$_\text{odd}$ and RLS$_\text{even}$ folds [see e.g.~profile of a four peak state in panel (c)]. Just as in the active PFC model, here, too, the asymmetric rung states are steady and remain at rest as $\alpha$ increases. Nevertheless, increasing nonreciprocity of the coupling does have a qualitative effect. Hopf bifurcations (diamond symbols) arise on all the branches shown in Fig.~\ref{fig:CH+CH_snaking} as $\alpha$ increases and the segments of stable steady states shrink with increasing activity~\cite{FrTh2021ima}. Within the snaking structure, Hopf bifurcations arise via Bogdanov-Takens bifurcations at the folds and also at the pitchfork bifurcations to the asymmetric states. Figure~\ref{fig:CH+CH_snaking}(d) shows an example of this behavior for $\alpha= 1.78$ and focuses on a small parameter region within the snaking structure where two successive Hopf bifurcations appear on the branch of symmetric states while one Hopf bifurcation appears on the branch of asymmetric states. These bifurcations change the linear stability behavior of the steady states in this parameter region. In panel~(d) the $+$ symbols indicate the number of unstable eigenvalues calculated via numerical stability analysis on the full domain, while the thin black lines emerging from these bifurcations represent the resulting branches of standing oscillations [cf. panels~(e)-(g)]. Note that panel (f) represents an in-phase oscillation of the two fronts while (g) shows a similar oscillation that is in anti-phase. Such oscillations are expected when an even parity state loses stability in a Hopf bifurcation; moreover, we expect that for wider even parity states the two Hopf bifurcations will approach one another, becoming almost degenerate. Panel~(e) shows an asymmetric though standing oscillation originating from the branch of asymmetric steady states. \New{In a generic system one would expect this state to drift as well as oscillate.}

Thus, here, an increase in activity does not lead to the onset of drift but is instead responsible for the presence of a pair of new branches of (unstable) time-periodic solutions near the left folds (as well as for asymmetric oscillations), thereby reducing the stability interval of the existing steady states. Since the upper Hopf bifurcation in (d) is subcritical, this bifurcation does not generate stable oscillations but leads instead to the elimination of a pair of peaks, one on either side of the structure, and a transition to a stable 7 peak state. In contrast, near the right folds Hopf bifurcations are only present when the RLS are short. Hopf bifurcations within the snaking structure have also been found in other systems (see e.g.~\cite{BuDa2012sjads}), but not on branches of asymmetric solutions. This may be because in our case the asymmetric states remain steady, while in generic active systems such states necessarily drift and a Hopf bifurcation from such states would lead to two-frequency localized states \cite{ORSS1983pd}.
	
Next, we examine how drift-transcritical bifurcations arise in this model. As for the active PFC model, we need to find a coexistence between asymmetric steady and drifting states so we can track these states. For this purpose we begin with the inversion-symmetric case $\bar \phi=\Delta \bar \phi = 0$ for which \eqref{eq:CHCH} is invariant with respect to $\mathcal{R}$ and $\mathcal{I}$, and choose a rather large $\alpha$, so that time-periodic behavior is preferred, and trace this back to drift-pitchfork bifurcations from symmetric \New{(even parity symmetry)} or antisymmetric \New{(odd parity symmetry)} steady states. We then gently break the inversion symmetry of the system so that the antisymmetric states lose their antisymmetry and become asymmetric, and ask whether these states remain at rest and the corresponding drift-pitchfork bifurcation becomes a transcritical one.
	
\begin{figure}[h!]
		\begin{minipage}[c]{\textwidth}
			\includegraphics[width=\textwidth]{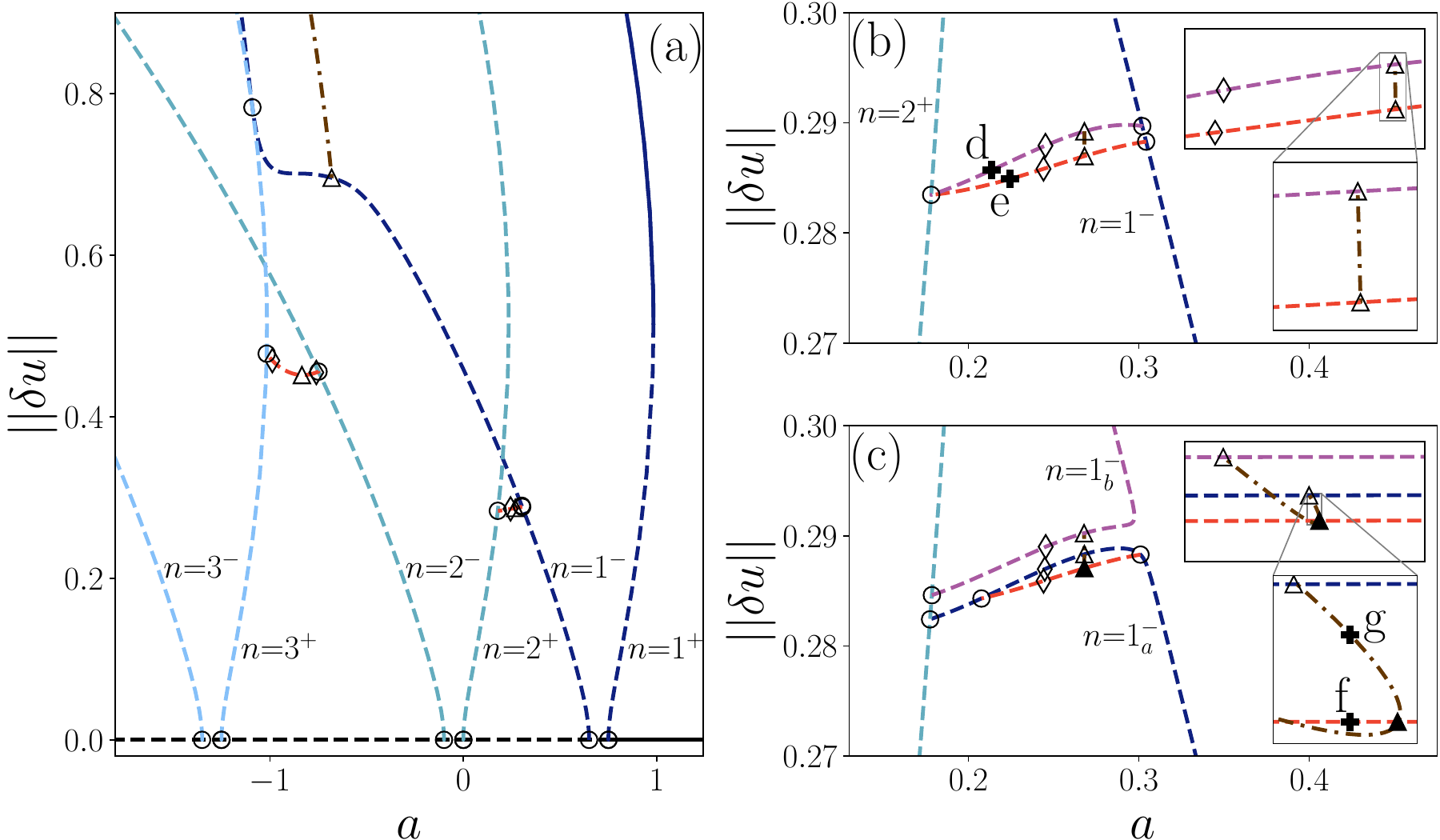}~\\
			\vspace{0.1cm}
			\includegraphics[width=\textwidth]{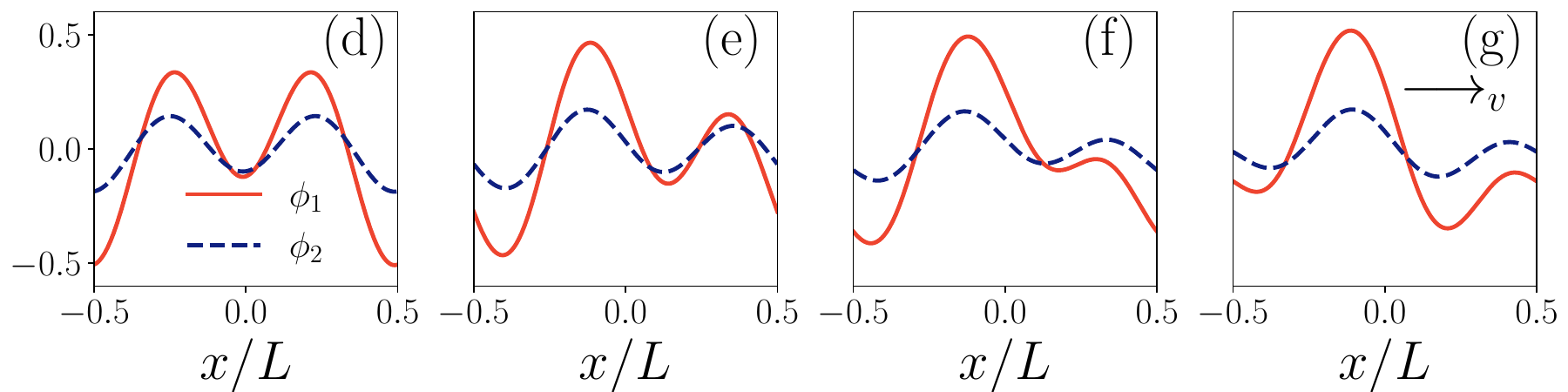}
		\end{minipage}
		\caption{(a) The L$_2$ norm as a function of the effective temperature $a$ for the two coupled Cahn-Hilliard equations~\eqref{eq:CHCH} in the inversion-symmetric case $\bar \phi = \Delta \bar \phi = 0$. Panel (b) highlights a parameter region of (a) where even [panel~(d)] and odd [panel~(e)] parity states connect the primary branches $1^-$ and $2^+$. Panel (c) shows the same parameter range as (b), but for the case when the inversion symmetry is weakly broken ($\bar \phi= 0.002$, $\Delta\bar \phi= 0$). Insets in~(b) and~(c) highlight the branch of drifting states [panel~(g)] that connect branches of steady parity-symmetric and asymmetric states [panel~(f)] via drift-pitchfork (empty triangles) and drift-transcritical (filled triangles) bifurcations, respectively. Other parameters are $\alpha = 1.65$, $\rho = 1.35$, $\kappa = 1$, $a_\Delta = -1.9$. The domain size is $L=4 \pi$. Line styles and symbols are as in Table~\ref{tab:instab}.}
		\label{fig:CH+CH_subcritical} 
\end{figure}	
	
Figure~\ref{fig:CH+CH_subcritical} presents the bifurcation diagram in the inversion-symmetric case. Here, the effective temperature $a$ is used as the main control parameter. For the given parameters, the model exhibits a large-scale instability, i.e., at the first primary bifurcation the fully phase-separated state ($n=1^+$) appears. We label each branch with a number indicating the spatial period of the corresponding solution and additionally an index $\pm$ indicating to which eigenvalue of the dispersion relation the associated primary bifurcation belongs. Interestingly, all $+$ branches arise subcritically, even though no quadratic (in general destabilizing) nonlinearities appear in the individual Eqs.~\eqref{eq:CHCH}. Nevertheless, as explained in more detail in Ref.~\cite{FrWT2021pre}, subcritical behavior can occur in a two-component system with nonreciprocal coupling even in this case. In the context of this work, we focus on the secondary and tertiary bifurcations and the corresponding branches that emerge. First there are two pairs of similar branches that connect the $1^-$ branch to the $2^+$ branch and the $2^-$ branch to the $3^+$ branch via pitchfork bifurcations. The former are also highlighted in a zoom [panel (b)] where we see that one drift-pitchfork (triangle symbol) and one Hopf bifurcation (diamond symbol) occur on each branch [see inset for magnification]. The corresponding states exhibit even [light purple line in (b), profiles in (d)] and odd [red line in (b), profiles in (e)] parity, respectively, and are therefore at rest (cf.~Sec.~\ref{sec:generic}). Two drift-pitchforks, one on each of the secondary branches, break the even or odd symmetry of these states, resulting in traveling asymmetric states [dashed-dotted brown line in (b), see inset for magnification], as one would expect for asymmetric states in nonvariational systems. Moreover, one finds that the resulting branch of asymmetric traveling states connects the two secondary branches of even and odd parity-symmetric steady states, cf.~Refs.~\cite{dang1986dss,CrKR1990pd}.

Next, we set $\bar \phi \neq 0$, thereby breaking the inversion symmetry. The corresponding bifurcation diagram is shown in Fig.~\ref{fig:CH+CH_subcritical}(c). The figure shows that the broken inversion symmetry has different consequences for the secondary bifurcations on the $1^-$ and $2^+$ branches. In particular, the upper pitchfork bifurcation on the $1^-$ branch that gives rise to the light purple branch in panel~(b) becomes imperfect in (c), resulting in a saddle-node bifurcation on the light purple branch $1_b^-$ in (c) and a single-valued branch $1_a^-$ [dark blue line in (c)]. Both $1_a^-$ and $1_b^-$ still connect to the $2^+$ branch (light blue line) but the termination points now differ. This is a consequence of the splitting of the degenerate pitchfork bifurcation on the $2^+$ branch in (b) into three separate pitchfork bifurcations, two of which remain on the $2^+$ branch while the third one takes place on the $1^-_a$ branch just prior to its termination on the $2^+$ branch.
        
The splitting of the branch of even parity states [light purple line in (b)] into two different branches that remain parity-symmetric in (c) also doubles the number of the corresponding drift-pitchfork and Hopf bifurcations [triangle and diamond symbols on the light purple and dark blue branches in (c), respectively]. Importantly, the branch of antisymmetric states in (b) is still present in (c), but now represents asymmetric, but still steady, states [red line in (c), profile in (f)] that arise in the two pitchfork bifurcations on the $1^-_a$ branch. The Hopf bifurcation on the red branch of (b) is also present in (c), and likewise for the drift bifurcation (filled triangle symbol) representing onset of motion where two branches of traveling asymmetric states emerge [see dashed dotted brown lines in insets in (c), profile in (g)]. However, this bifurcation is now a transcritical drift bifurcation, and one of the emerging branches exhibits a saddle-node bifurcation [see bottom inset in (c)] before each branch of the resulting traveling states connects to one of the two now distinct branches of symmetric steady states via a drift-pitchfork bifurcation (triangle symbols). That is, owing to the broken inversion symmetry, steady antisymmetric states become steady asymmetric states while the drift-pitchfork bifurcation splits into a drift-transcritical and a saddle-node bifurcation.
        
With increasing inversion symmetry breaking, i.e., for increasing $\bar \phi$, the branch of steady asymmetric states [red branch in (c)] shrinks until the corresponding pitchfork bifurcations annihilate and the drift-transcritical bifurcation vanishes.

\subsection{Coupled Cahn-Hilliard and Swift-Hohenberg equations}\label{sec:CH+SH}

Next, we examine a two-variable model that couples the Cahn-Hilliard equation to the nonconserved Swift-Hohenberg equation. The classical (nonconserved) Swift-Hohenberg equation was originally introduced to describe pattern selection in Rayleigh-B\'enard convection \cite{SwHo1977pra}. The equation was subsequently identified as a simple but useful model capable of providing a qualitative understanding of various pattern-forming systems. Applications run from reaction-diffusion systems \cite{GuOS1994pre} to laser physics \cite{LeMN1994prl,TlidiPRL1994} and fluid dynamics \cite{ScGB2010prl,MBAK2013jfm}, and even to growth processes in ecosystems \cite{Mero2012em}. The bistable SH equation captures essential properties such as wavelength selection and the properties of spatially localized states such as fronts and pulses and their pinning and depinning. In particular, there have been numerous studies of the properties of localized structures and of the so-called homoclinic snaking these structures exhibit in both one and two dimensions~\cite{WoCh1999pd,LSAC2008sjads,BuKn2007c}.
    
The Cahn-Hilliard and Swift-Hohenberg models are minimal models for two different stationary instabilities of the trivial homogeneous state. The Swift-Hohenberg model describes a nonconserved quantity exhibiting a small-scale instability with a finite critical wavenumber which determines the emerging pattern. In contrast, the Cahn-Hilliard equation models a large-scale instability of a conserved quantity where demixing results in a phase-separated state. A model that couples the Swift-Hohenberg equation to the Cahn-Hilliard equation thus describes the interaction between pattern formation and phase separation. Near the codimension-2 point, where both instabilities occur with comparable growth rates the pattern-forming process competes with demixing as observed, e.g.,~in Marangoni convection in thin liquid films \cite{VSSM1997jfm}, resulting in states that may consist of different spatial scales and exhibit additional secondary instabilities~\cite{Metz2001}.

Here we employ a linear interaction between the two fields. Again, it can be separated into reciprocal and nonreciprocal parts controlled by the parameters $\rho$ and $\alpha$, respectively. The coupled model reads

\begin{alignat}{1}
\begin{aligned}
\qquad \frac{\partial\phi_1}{\partial t}  &= \frac{\partial^2}{\partial x^2} \left(- \kappa\frac{\partial^2\phi_1 }{\partial x^2} +f_1'(\phi_1) + \left(\rho + \alpha\right) \phi_2 \right)\,,\\
\qquad \frac{\partial\phi_2}{\partial t}  &=  -\left[ \left(q^4 -r  + 2 q^2 \frac{\partial^2}{\partial x^2}  + \frac{\partial^4}{\partial x^4} \right) \phi_2 + f_2'(\phi_2) + \left(\rho - \alpha\right) \phi_1 \right]
\end{aligned}
\label{eq:CHSH}
\end{alignat}
with
\begin{alignat}{1}
\begin{aligned}
f_1=&  \frac{a}{2} \phi_1^2  + \frac{1}{4} \phi_1^4\,,\\
  f_2=& - \frac{\delta}{3} \phi_2^3  + \frac{1}{4} \phi_2^4\,.
\end{aligned}
\label{eq:CHSH_ff}
\end{alignat}
Since the individual equations both obey gradient dynamics, the coupled model becomes active, i.e., nonvariational, only when $|\alpha|>|\rho|$. The dynamics of $\phi_1$ remain mass-conserving, and at all times
\begin{alignat}{1}
\begin{aligned}
\frac{1}{L} \int_{-\frac{L}{2}}^\frac{L}{2} \textrm{d}x \,\phi_1 = &\bar \phi_1\,.
\end{aligned}
\label{eq:CHSH_int}
\end{alignat}

\begin{figure}[h!]
\includegraphics[width=\textwidth]{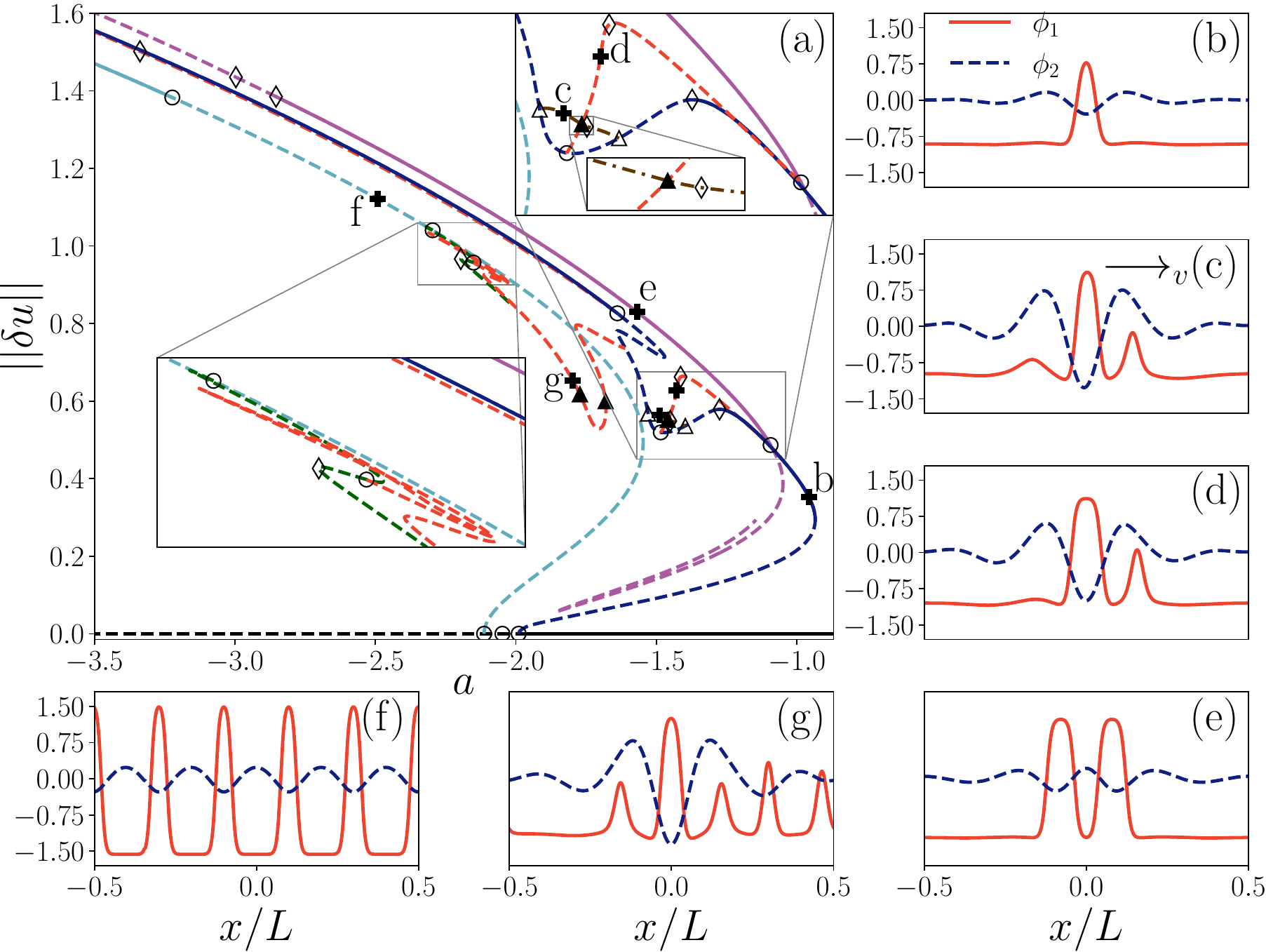}
\caption{(a) The L$_2$ norm \eqref{eq:norm}, with $u_i=\phi_i$, as a function of $a$ for various states described by the nonvariationally coupled CH and SH equations~\eqref{eq:CHSH} at reciprocal and nonreciprocal interaction parameters $\rho=0.2$ and $\alpha=0.5$, respectively. The homogeneous state (black line) corresponds to $(\bar \phi_1, \bar \phi_2) \approx (-0.8,-0.067)$ where $\bar \phi_1$ is also the mean value of $\phi_1$ for all the other states. Branches of steady asymmetric states are indicated by red lines. Line styles and symbols are as in Table~\ref{tab:instab}. Panels~(b)-(e) show sample profiles at locations indicated by crosses in panel (a). The remaining parameters are $\kappa=0.04$, $r=-0.5$, $\delta = 3.2$ and $q=1.3$. The domain size is $L=5 \pi$.}
\label{fig:CHSH_bifdiag} 
\end{figure} 

Figure~\ref{fig:CHSH_bifdiag} presents a bifurcation diagram and selected profiles of various steady and drifting states, including steady asymmetric states (red lines). The parameter $a$ is used as the control parameter, the characteristic wavenumber of the Swift-Hohenberg equation is set to $q=1.3$; with $\rho=0.2<\alpha=0.5$ the model \eqref{eq:CHSH} is nonvariational, i.e., we expect time-periodic behavior.

For a given mean density $\bar \phi_1$ the homogeneous state is $(\bar \phi_1, \bar \phi_2)$ where $\bar \phi_2$ solves $(q^4-r) \bar \phi_2 + f_2'(\bar \phi_2) + \left(\rho - \alpha\right) \bar \phi_1=0$. With $\bar \phi_1=0.8$ this equation gives $\bar \phi_2 \approx -0.067$ and the resulting homogeneous state (black line) is stable for $a \gtrsim -1.99 $, where it becomes unstable with respect to a large-scale instability inherited from the Cahn-Hilliard equation. This instability leads to a subcritical branch of period-one solutions (dark blue line) that localize and gain stability in a fold at $a \approx -0.93$. The resulting stable solutions are characterized by a pronounced peak in the $\phi_1$-field at a location where $\phi_2$ is a minimum [cf. panel~(b)]. Note that due to an effective run-and-chase interaction ($\rho + \alpha>0$, $\rho-\alpha<0$), $\phi_2$ [$\phi_1$] favors alignment [anti-alignment] with $\phi_1$ [$\phi_2$]; since $|\bar \phi_1| > |\bar \phi_2|$, $\phi_1$ is dominant and it is therefore reasonable that anti-alignment is preferred. For decreasing $a$ these one-peak solutions lose stability in a Hopf bifurcation (empty diamond) at $a \approx -1.28$, followed by drift-pitchfork (empty triangle) and pitchfork (empty circle) bifurcations at $a\approx -1.40$ and $a \approx -1.48$, respectively; these states are subsequently restabilized via a second drift-pitchfork bifurcation, followed by a pair of fold bifurcations and a further pitchfork bifurcation at $a \approx -1.64$. With further decrease of $a$ the one-peak states remain stable but acquire a pair of side peaks, i.e., they become symmetric three-peak states (not shown). Having described the branch of one-peak solution we now return to the aforementioned two drift-pitchfork bifurcations and the pitchfork bifurcation between them [see upper right inset in panel~(a)]. The diagram shows that the drift-pitchfork bifurcations are connected by a branch of drifting asymmetric states [dashed dotted brown line, panel~(c)]. In contrast, the pitchfork bifurcation leads to a branch of {\it steady} asymmetric states [red line, panel~(d)] which collides with the branch of drifting states in a drift-transcritical bifurcation (filled triangle, see smaller inset for magnification). Both branches of asymmetric states exhibit Hopf bifurcations which lead to modulated waves and standing asymmetric oscillations that emerge from the drifting and steady asymmetric states, respectively. The red branch finally connects via a pitchfork bifurcation to a symmetric two-peak state [light purple line, panel~(e)], thereby stabilizing the two-peak states for lower values of $a$ before they lose stability again via two successive Hopf bifurcations. Tracking the light purple branch back, i.e., for increasing $a$, we see that it exhibits a fold but does not connect to the homogeneous state. Instead it turns around at small amplitude in a sharp fold. The bifurcation behavior that follows resembles that of the one-peak states, occurs in a similar parameter range, but is much more complex. We omit these details. Other two-peak states emerge from the second primary bifurcation of the trivial state, but are also omitted here.

Next, we consider the bifurcation behavior connected to the branch emerging in the third primary bifurcation from the homogeneous state. The resulting branch consists of five equispaced peaks [light blue line, panel~(f)] instead of the three-peak states one might expect for a large-scale instability. This is due to the chosen parameters that bring the system close to a codimension-2 point where both large- and small-scale instability occur simultaneously. The light blue branch also emerges subcritically this time with five unstable eigenvalues. The branch gains linear stability through a fold and two degenerate, i.e., double, pitchfork bifurcations, both of which ultimately lead to asymmetric states. The behavior near the first one, at $a \approx -2.30$, is enlarged in the corresponding inset. The pitchfork bifurcation of the five-peak periodic state generates a subcritical branch bifurcating towards decreasing $a$ values [dark green line] consisting of a reflection-symmetric five-peak state with unequal spacing. The dark green branch turns back at the left fold and back again at the next fold on the right, followed by a pitchfork and a Hopf bifurcation. Further complex bifurcation behavior occurs beyond these points and is omitted. At the aforementioned pitchfork bifurcation the reflection symmetry of the solution is broken and a new branch of steady asymmetric states emerges [red line, panel~(g)]. This branch also snakes back and forth through four folds before continuing to larger $a$ values. Near the fifth fold where the branch turns back again two drift-transcritical bifurcations (filled triangles) occur where the steady asymmetric states connect to drifting asymmetric states. Beyond the sixth fold the asymmetric steady states terminate in a pitchfork bifurcation on symmetric states; the corresponding branch is omitted (it is not the dark blue branch).

In addition, we have found various steady asymmetric states that show asymmetric peak-to-peak separations, e.g.~at the second pitchfork bifurcation of the dark blue branch where the three-peak state regains stability such a steady asymmetric state occurs (red line) where one of the outer peak becomes broader, the other narrower. Since this asymmetric state has a similar norm as its symmetric pendant, both the symmetric and asymmetric branches are almost identical.

Much of the structure just described owes its existence to the presence of a subcritical Turing bifurcation (where the light blue branch emerges) as described further in \cite{KnYo2021ijam} and a strong quadratic nonlinearity ($\delta =3.2$) that creates a coupling of neutral and pattern mode \cite{MaCo2000non}. Evidently the fact that the Turing bifurcation is preceded by additional primary bifurcations to large-scale states is a consequence of the coupling to the Cahn-Hilliard equation.
\New{However, as in the preceding examples, the presence of asymmetric steady states indicates that the model has a hidden nongeneric structure. Here, all the asymmetric steady states are linearly unstable and we would not expect them to be readily detectable in experiments. In contrast, the model studied next does exhibit linearly stable steady asymmetric states.}

\subsection{\Resub{Two-species active PFC model}}\label{sec:a+pPFC}

In this section we describe the properties of a three-variable model, comprising a passive PFC model coupled to an active one \cite{VHKW2022msmse}:
\begin{alignat}{1}
\begin{aligned}
\qquad \partial_{t}\phi_1 &= \partial_{xx}\left\{\left[\epsilon+\left(1+\partial_x^{2}\right)^{2}\right]\phi_1+\left(\bar{\phi}_1+\phi_1\right)^{3} + c \phi_2 \right\}-v_{0}\partial_x P\,, \\
\qquad \partial_{t} P &= \partial_{xx}\left(c_1 P + c_2 P^3\right) - D_{\mathrm{r}}(c_1 P +  c_2 P^3)-v_{0} \partial_x \phi_1\,,~\\
\qquad \partial_{t}\phi_2 &=\partial_{xx}\left\{\left[\epsilon+\left(1+\partial_x^{2}\right)^{2}\right]\phi_2+\left(\bar{\phi}_2+\phi_2\right)^{3} + c \phi_1 \right\} \,.
\end{aligned}
\label{eq:apfc2}
\end{alignat}
The two densities $\phi_1$, $\phi_2$ are coupled via the parameter $c$. The latter obeys gradient dynamics, while the former is nongradient when the activity parameter $v_0\ne 0$. As for the \Resub{one-species} active PFC model in Sec.~\ref{sec:aPFCModel}, Eqs.~\eqref{eq:apfc2} are parity-symmetric in the sense of being invariant w.r.t.~$\mathcal{R}: \left(x, \phi_1, P,\phi_2 \right) \mapsto \left(-x, \phi_1, -P,\phi_2\right)$. \Resub{The model~\eqref{eq:apfc2} may be systematically derived from microscopic theory: it corresponds to \New{an approximation to} a more general PFC model that is itself systematically derived from dynamical density functional theory~\cite{VHKW2022msmse}.} \New{In particular, it describes the interaction of passive and active Brownian particles.}

\begin{figure}[h!]
	\centering
	\includegraphics[width=\textwidth]{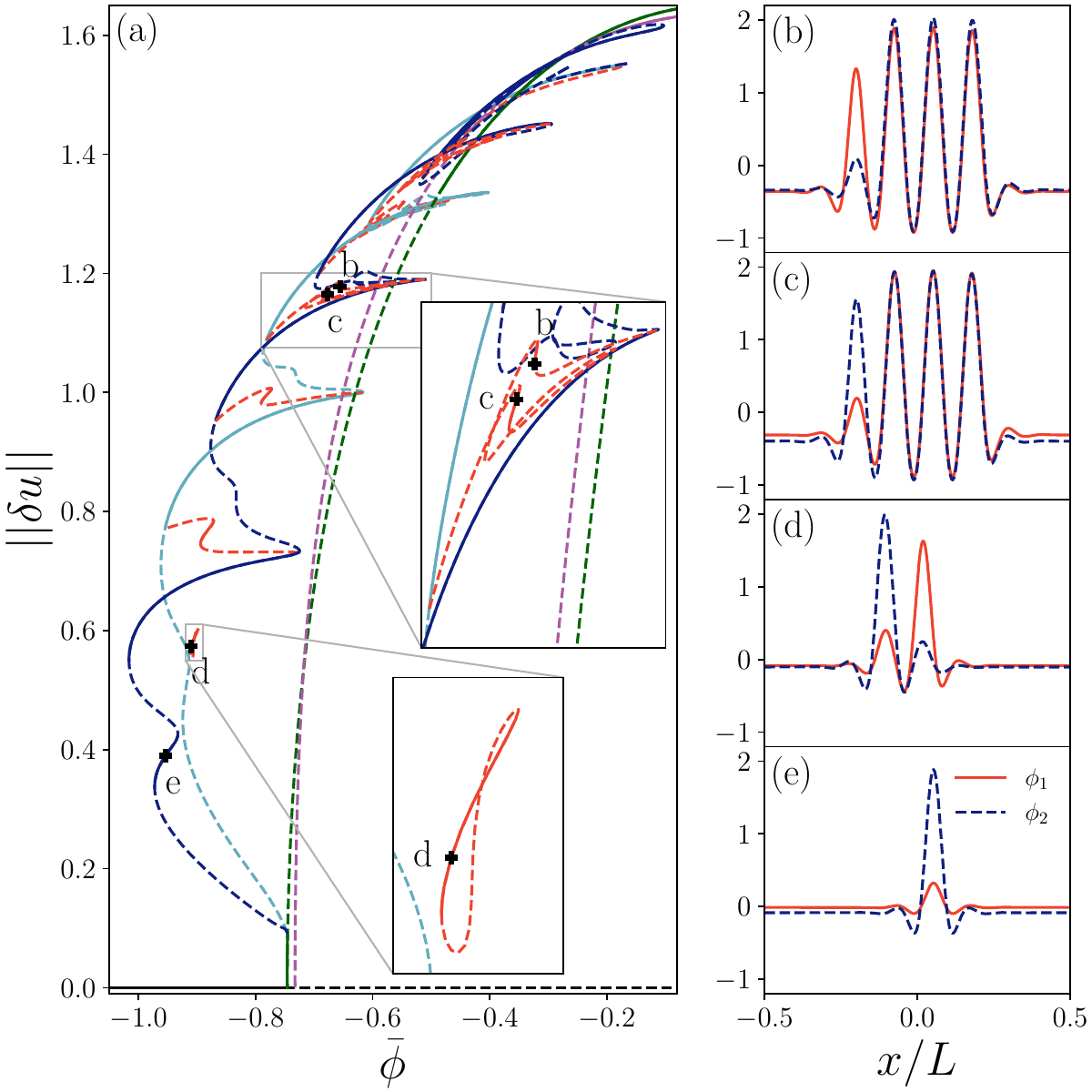}
	\caption{Bifurcation diagram (left) and solution profiles (right) for the \Resub{two-species active PFC model}~\eqref{eq:apfc2}. Panel~(a) shows the L$_2$ norm~\eqref{eq:norm} as a function of the conserved mean density $ \bar{\phi}\equiv\bar{\phi}_1=\bar{\phi}_2 $.  The homogeneous state (black) is destabilized in a primary bifurcation at $ \bar{\phi} \approx -0.75 $. From this bifurcation, the branch of periodic states (dark green) with $ n = 8 $ peaks emerges supercritically. The branch of periodic states is destabilized by a secondary bifurcation to RLS$_\text{odd}$ (dark blue) and RLS$_\text{even}$ (light blue). The branches of localized states are connected by branches (red) of resting, asymmetric localized states. Panels~(b)-(e) show stable resting states at the locations indicated by the corresponding labels in the main panel~(a). The solution profiles show the fields $ \phi_1 $ (solid red) and $ \phi_2 $ (dashed dark blue). Line styles are as in Table~\ref{tab:instab}; here, bifurcations are not indicated by corresponding symbols. The remaining parameters are $ r = -1.5$, $q_1 = q_2 = 1$, $c=-0.2$, $v_0=0.1$, $c_1=0.1$, $c_2=0$, $D_r=0.5$. The domain size is $ L = 16\pi $.}
	\label{fig:apfc2_snaking_phi0j}
\end{figure}

\begin{figure}[h!]
	\centering
	\includegraphics[width=0.8\textwidth]{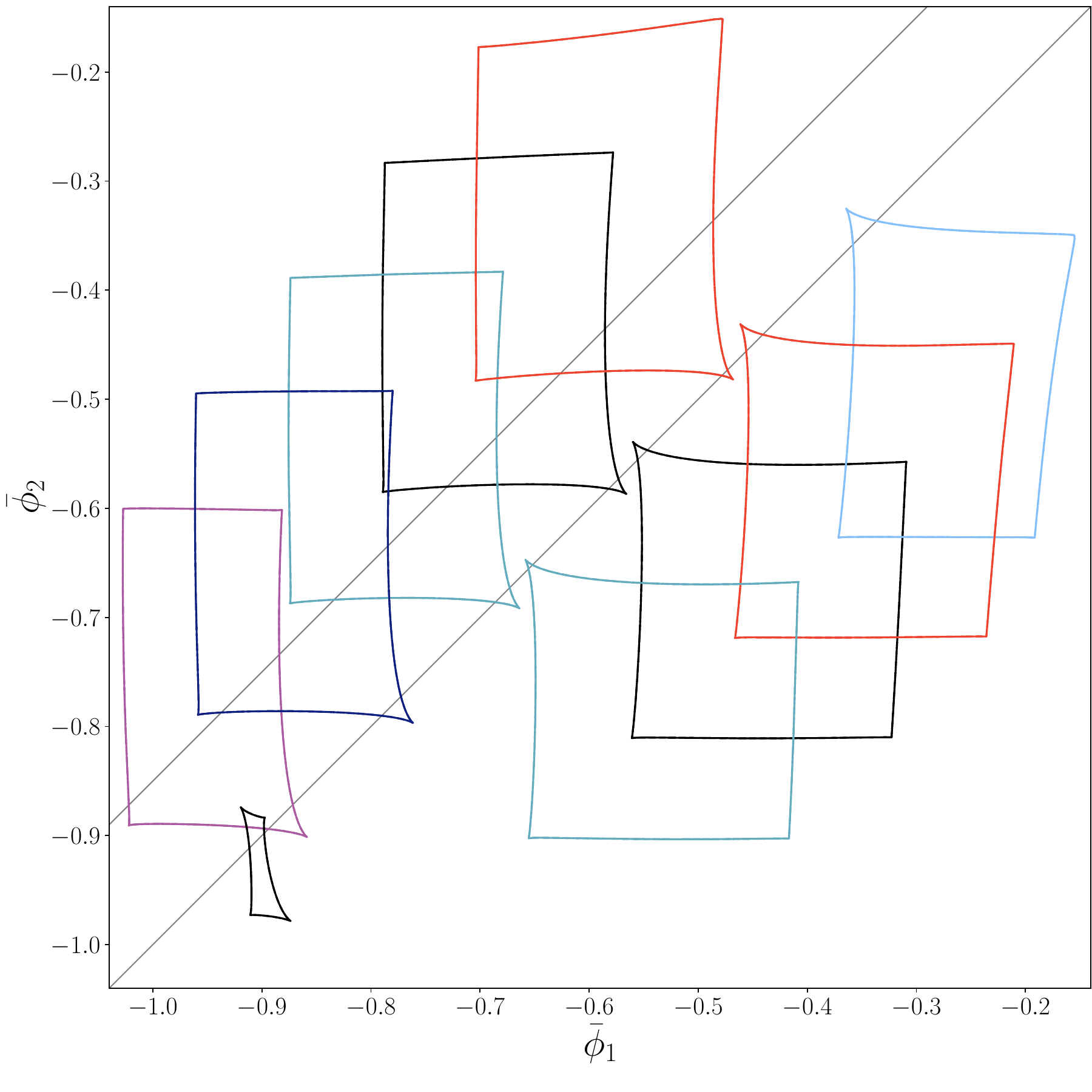}
	\caption{Two-parameter continuation of the saddle-node bifurcations that frame the stable, resting, asymmetric localized states in Fig.~\ref{fig:apfc2_snaking_phi0j} in the $(\bar{\phi}_1,\bar{\phi}_2)$ plane. Paths of a given color are from the same ladder branch. The lower [upper] thin gray diagonal line shows the continuation path in Fig.~\ref{fig:apfc2_snaking_phi0j} [Fig.~\ref{fig:apfc2_snaking_shiftphi02_0.15_phi0j}]. Stable, resting, asymmetric localized states are present between each pair of same-color lines cut by the diagonal lines. The remaining parameters are the same as in Fig.~\ref{fig:apfc2_snaking_phi0j}.}
	\label{fig:apfc2_asymm_folds}
\end{figure}

\begin{figure}[h!]
	\centering
	\includegraphics[width=\textwidth]{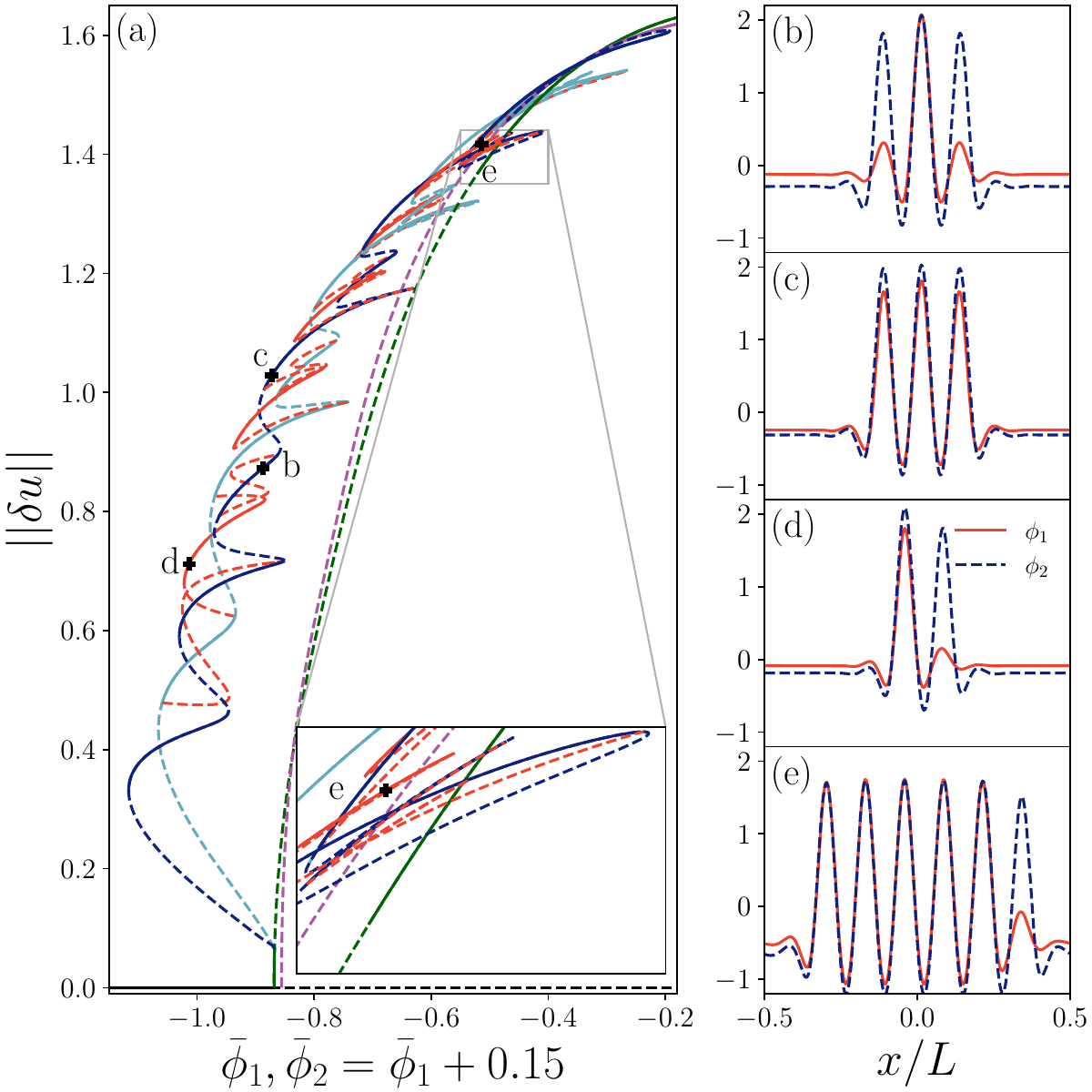}
	\caption{(a) One-parameter continuation along $ \bar{\phi}_1 $ when $ \bar{\phi}_2 = \bar{\phi}_1 + 0.15 $ following the upper gray line in Fig.~\ref{fig:apfc2_asymm_folds}. The right panels show solution profiles at the locations labeled in the main panel~(a). Panels (b) and (c) show stable, symmetric RLS$_\text{odd}$ states. Panels (d) and (e) show two stable, resting, asymmetric states. Line styles, norm and the remaining parameters are the same as in Fig.~\ref{fig:apfc2_snaking_phi0j}.}
	\label{fig:apfc2_snaking_shiftphi02_0.15_phi0j}
\end{figure}

Figure \ref{fig:apfc2_snaking_phi0j}(a) summarizes the steady states of Eqs.~\eqref{eq:apfc2} at low activity, $ v_0 = 0.1 $, as a function of the conserved mean density $\bar{\phi}\equiv \bar{\phi}_1 = \bar{\phi}_2 $. The homogeneous state is stable for low $ \bar{\phi}$ but loses stability in a pitchfork bifurcation at $ \bar{\phi}\approx -0.75 $. From this bifurcation a branch of periodic states with $ n=8 $ emerges supercritically. Shortly thereafter the periodic states are destabilized in another pitchfork bifurcation, from which two branches of localized states, with odd and even numbers of peaks, emerge. These exhibit homoclinic snaking with new peaks added near the left folds. Near these the rung branches of unstable, resting, asymmetric states emerge in pitchfork bifurcations, as in the previous examples. However, at this choice of parameters, there is at least one segment of {\it stable}, resting, asymmetric states on each rung branch generated via a pair of folds. Similar restabilization has been observed for the two-dimensional Swift-Hohenberg equation \cite{ALBK2010sjads}. Panels~(b) and (c) of Fig.~\ref{fig:apfc2_snaking_phi0j} show two distinct types of stable, resting, asymmetric states. In panel~(b) [(c)] there is an additional $ \phi_1 $-[$ \phi_2 $-]peak at the left interface of a three-peak localized state. Both states are located on the same rung branch, highlighted in the magnified inset in the main panel (a). Panel~(d) shows a third kind of resting, asymmetric state: here, in contrast to 'normal' localized states, the peaks of the two fields no longer coincide but are a wavelength apart. The peaks in each field are locked to the oscillatory tail of the other field. This state forms an isola with stable states between two saddle-node bifurcations. The most narrow stable localized state is found on the RLS$_\text{odd}$ branch [panel~(e)] and consists of a single large amplitude peak of $ \phi_2 $ and a single low amplitude peak of $ \phi_1 $, both at the same location.

The saddle-node bifurcations that frame each stable segment on the rung branches can be tracked using two-parameter continuation (Fig.~\ref{fig:apfc2_asymm_folds}). Figure \ref{fig:apfc2_snaking_shiftphi02_0.15_phi0j} shows that stable, resting, asymmetric localized states are present when $v_0\ne0$ and $\bar{\phi}_2\ne\bar{\phi}_1$; the latter condition results in the splitting of the rung states.

\subsection{Reaction-diffusion system}\label{sec:rd-system}

Having shown that nongeneric steady asymmetric states are features of several active matter models we now briefly indicate their existence in a standard reaction-diffusion model, namely, the FitzHugh--Nagumo (FHN) model. The model consists of two nonconserved variables and their dynamics determined by diffusion and local nonlinear kinetics. Without diffusion the coupled ordinary differential equations were originally used to describe nerve impulses~\cite{Fitz1961bj,NaAY1962ire}. The corresponding reaction-diffusion model describes an activator-inhibitor system which can be used to study, e.g.,~collisions of nonlinear waves in excitable media~\cite{ArCK2000jtb}, self-sustained oscillations in semiconductor amplifiers~\cite{BPGT2003pre}, and diffusion-driven (Turing) instabilities resulting in the emergence of self-organized pattern formation~\cite{Turi1990bmb,NaMi2010np}.

The FHN model is neither variational, nor mass-conserving. Despite this, we argue that some versions of the model are also nongeneric, in the same sense as the models considered hitherto.

We consider here the two-variable model 
\begin{alignat}{1}
\begin{aligned}
\qquad \partial_t u_1 =  D \partial_{xx} u_1 + f(u_1) - u_2 + \kappa + \varepsilon u_2^3\,,~\\
\qquad \delta \partial_t u_2 = \partial_{xx}u_2 - u_2 +u_1\,,
\end{aligned}\label{eq:rd}
\end{alignat}
where $f(u_1) = \lambda u_1 - u_1^3$ as in Ref.~\cite{ScBP1995pd}. In Eqs.~\eqref{eq:rd} $u_1$ and $u_2$ denote the activator and inhibitor, respectively, and $D$ is the ratio of their diffusion constants; $\kappa$ is a source term for the activator and is used as the control parameter. The time scale ratio $\delta$ drops out when considering steady states. When $\varepsilon=0$ the local kinetics are of FitzHugh--Nagumo type; for $\varepsilon\neq0$ the additional term $\varepsilon u_2^3$ represents a generalization that corresponds to an amplification ($\varepsilon>0$) or weakening ($\varepsilon<0$) of the inhibition for large $u_2$ values. \New{For our purposes the additional term is key because the resting asymmetric localized states that are present for $\varepsilon = 0$ begin to drift whenever $\varepsilon\neq 0$. In other words, the additional term destroys the nongeneric nature of the model.}

\begin{figure}[h!]
	\centering
	\includegraphics[width=\textwidth]{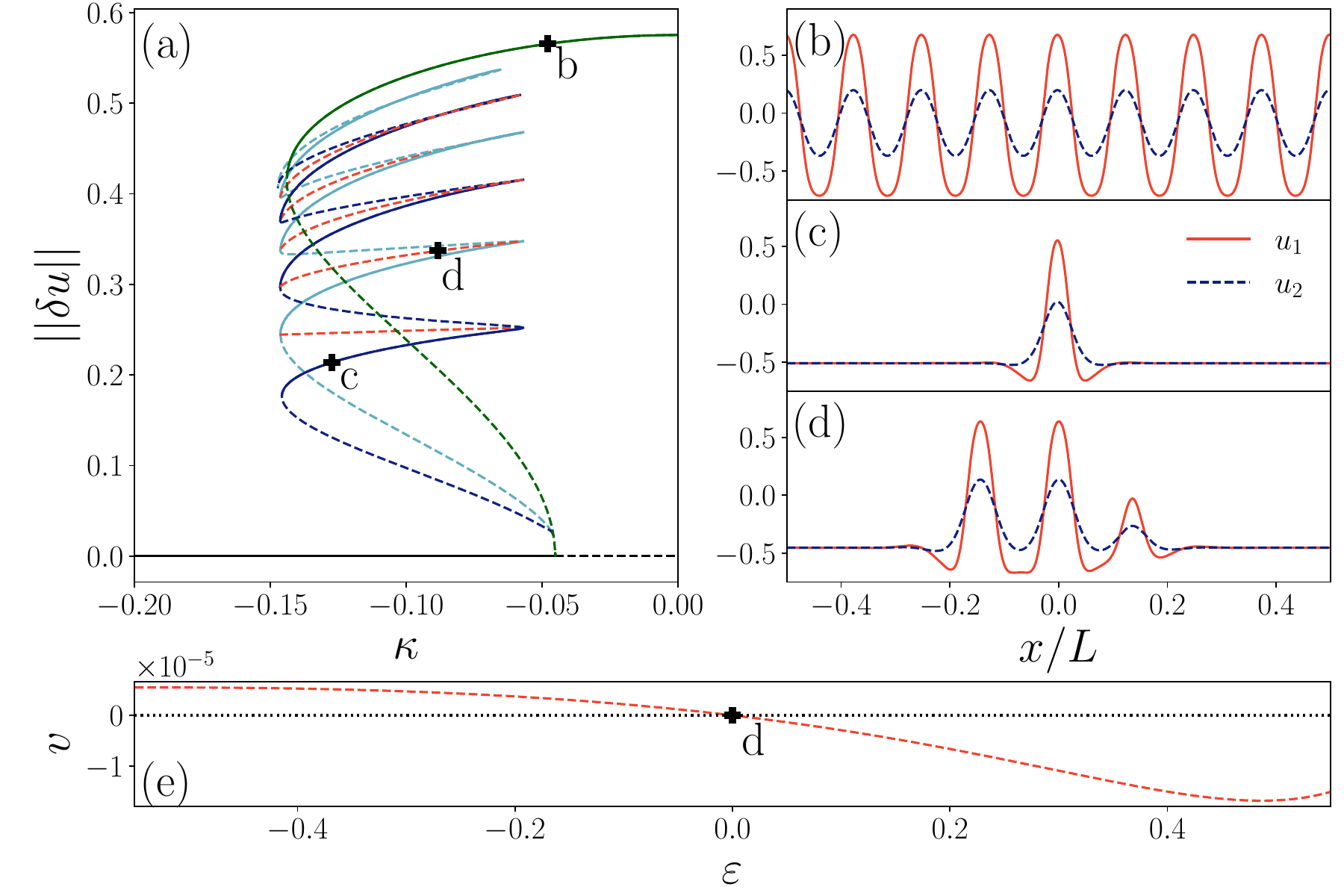}
	\caption{The reaction-diffusion system \eqref{eq:rd} exhibits standard homoclinic snaking, shown in panel~(a) in terms of the L$_2$ norm~\eqref{eq:norm} as a function of the parameter $\kappa$. The homogeneous state (black) is destabilized in a primary bifurcation at $ \kappa \approx -0.045$. From this bifurcation, a branch of periodic states (dark green) with $ n = 8 $ peaks, shown in panel~(b), emerges subcritically. The branch of periodic states is further destabilized in a secondary bifurcation where branches of localized states with an odd (dark blue) and an even (light blue) number of peaks emerge. Panel~(c) shows an example of the former, with $ u_1 $ shown in solid red and $ u_2 $ in dashed dark blue.  The branches of localized states are connected by branches (red) of unstable, resting, asymmetric localized states, shown in panel~(d). The resting asymmetric states such as (d) start moving as soon as $\varepsilon\ne0$, as shown in panel~(e). The locations of the solution profiles (b)-(d) are indicated in panel~(a). Line styles are as in Table~\ref{tab:instab}; here, bifurcations are not indicated by corresponding symbols. The remaining parameters are $ D = 0.14 $, $ \lambda=1.01 $, $ \varepsilon = 0 $, $ \delta = 0.3 $. The domain size is $ L = 40 $. }
	\label{fig:rd_snaking}
\end{figure}

The equations are invariant w.r.t. $\mathcal{R}: x \mapsto -x$ only. As a result the parity-symmetric spatially localized states present in this model are again of two types, RLS$_\text{odd}$ and RLS$_\text{even}$, with odd (dark blue) and even (light blue) number of peaks, respectively, but this time organized within a standard snaking structure with no tilt. The resulting bifurcation diagram is shown in Fig.~\ref{fig:rd_snaking} for $\varepsilon=0$. In this case the presence of LS requires bistability between the trivial and the periodic states. Moreover, for this parameter value, the system also exhibits resting, linearly unstable asymmetric rung states (red), despite its nonvariational structure. Furthermore, these same states begin to drift as soon as $\varepsilon \neq 0 $ as depicted in panel~(e).

In the following section we analyze the nongeneric behavior exhibited by the five explicit models described in this section and explain its origin.

\section{When do steady asymmetric states exist?}
\label{sec:existence}
\subsection{\New{Spurious gradient dynamics}}

We now propose a general form of a family of multicomponent models that allows for resting asymmetric states even when \Resub{the model} is nonvariational, i.e., when the system is active. These multicomponent models typically have a ``dead limit'', i.e., a limiting case exists in which the system is variational. We show that within the proposed model family any resting asymmetric state that exists at a specific parameter value is part of a whole branch of such states. In particular, any steady asymmetric state that exists in the variational limit continues to do so as a resting state in the nonvariational case, at least over a finite parameter range. 

We assume a general $M$-component order parameter field $\vecg u=\left\{ u_i\right\}$ that consists of $M_1$ nonconserved components $n_i$ and $M_2$ conserved components $c_i$. We assume that the dynamics of the $i$-th component can be written as
\begin{align} \label{eq:stst_general0}
\partial_t u_i = 
\left\{
\begin{array}{l}
\partial_t n_i = -\left(\dfrac{\delta \mathcal{G}}{\delta n_i} +  \alpha_i \dfrac{\delta \mathcal{G}_C}{\delta n_i}\right)~\quad \text{for}\,\, i=1,\dots,M_1\,,\\
\partial_t c_i = \partial_{xx} \left(\dfrac{\delta \mathcal{G}}{\delta c_i} +  \alpha_i \dfrac{\delta \mathcal{G}_C}{\delta c_i} \right)~\quad  \text{for}\,\, i=M_1+1,\dots,M_1+M_2,
\end{array}
\right.
\end{align}
where $\mathcal{G}$ and $\mathcal{G}_C$ are two functionals containing self-interaction and coupling terms,
respectively. Furthermore, we introduce the operator $\mathcal{D}_{xx}^{(i)}$ to avoid an explicit distinction between conserved and nonconserved fields. For this purpose we simply define $\mathcal{D}_{xx}^{(i)}= -1$ if $i\le M_1$ and $\mathcal{D}_{xx}^{(i)}= \partial_{xx}$ if $i> M_1$. Equation~\eqref{eq:stst_general0} then takes the compact form
\begin{equation}\label{eq:stst_general}
\partial_t u_i =\mathcal{D}_{xx}^{(i)} \left( \frac{\delta \mathcal{G}}{\delta u_i} +  \alpha_i \frac{\delta \mathcal{G}_C}{\delta u_i}\right)\,.
\end{equation}
Here we only consider steady states. Because in this formulation mobilities do not alter the steady states of the system, we do not include them.
Since $\mathcal{G}$ only consists of self-interaction terms it can be written as a sum of independent contributions, i.e.
\begin{align}
\mathcal{G}[\vecg u]= \sum_i \mathcal{G}_i [u_i]\,.
\end{align}
If $\alpha_i=1 ~ \forall i$ we obtain the variational case with the  Lyapunov functional $\mathcal{F}= \mathcal{G} + \mathcal{G}_C$.
More precisely, even for arbitrary but purely positive entries, i.e., $\alpha_i>0 $, the system corresponds to a variational model since we can rewrite Eq.~\eqref{eq:stst_general} as
\New{
  \begin{align} \label{eq:nonconserved_rewritten}
\partial_t u_i = \alpha_i \mathcal{D}_{xx}^{(i)} \frac{\delta \mathcal{F}}{\delta u_i},
\end{align}
  with the Lyapunov functional $ \mathcal{F} =   \sum_i \alpha_i^{-1}   \mathcal{G}_i [u_i] +  \mathcal{G}_C $ and prefactors $\alpha_i$ that act as an effective (diagonal and positive definite) mobility matrix.} Note that for purely negative values of $\alpha_i$ one can incorporate the negative sign into the coupling terms $\dfrac{\delta \mathcal{G}_C}{\delta u_i}$, i.e., the system is still variational. In contrast, Eq.~\eqref{eq:stst_general} becomes nonvariational if the set of $\alpha_i$ contains both positive and negative values. This is because in this case the functional $\mathcal{F}$ is no longer bounded from below \textit{and} the effective mobility matrix is indefinite. \New{In particular, its eigenvectors corresponding to negative eigenvalues represent energy-increasing flow, in contrast to eigenvectors with positive eigenvalues that indicate the common energy-decreasing flow. The interplay of these two motions in the potential landscape, i.e., the indefinite property of the effective mobility matrix, lies at the heart of possible time-periodic behavior since the system is no longer required to evolve into a steady state, thereby reflecting its active character. Despite this, Eq.~\eqref{eq:nonconserved_rewritten} is still in gradient dynamics form. In the following we refer to this structure as \textit{spurious gradient dynamics}.}

In the following we take Eq.~\eqref{eq:stst_general} as given and consider steady states $\vecg{u}^{(0)}$ for which we replace the left hand side by zero. We integrate the steady state equation for the conserved fields once and assume that the integration constant, the flux, is zero.\footnote{The active PFC model is an example where the flux is not zero. We show in Sec.~\ref{sec:apfc} why the results of the present section still apply.} Evaluating the result and combining it with the spatial derivative of the steady state equations of the nonconserved fields we obtain
\begin{align}\label{eq:Ldxu0}
0=\sum_j \mathcal{L}_{ij} \partial_x u^{(0)}_j,
\end{align}
where $\underline{\mathcal{L}}$ is the operator with components
\begin{equation}\label{eq:L}
\mathcal{L}_{ij} \equiv \frac{\delta^2 \mathcal{G}}{\delta u_i \delta u_j}+ \alpha_i \frac{\delta^2 \mathcal{G}_C}{\delta u_i \delta u_j} =\frac{\delta^2 \mathcal{G}_i}{\delta u_i \delta u_i} \delta_{ij}+ \alpha_i \frac{\delta^2 \mathcal{G}_C}{\delta u_i \delta u_j} = \mathcal{L}_{ii}^G \delta_{ij}+ \alpha_i  \mathcal{L}_{ij}^{G_C}\,.
\end{equation}
Note that $\underline{\mathcal{L}}$ is evaluated at the steady state $\vecg u= \vecg{u}^{(0)}$.

\subsection{Existence of asymmetric steady states} \label{sec:deriv}
We proceed as follows. We assume that there exists a steady asymmetric state for a given choice of $\alpha_i=\alpha_i^{(0)}$. A natural example would be the asymmetric states that exist in the variational limit. We then consider infinitesimal but arbitrary changes $\delta \alpha_i$ in the parameters and show that the resulting slightly modified, but still asymmetric, state remains steady.

\paragraph*{Step 1: Derivation of the solvability condition}~\\
We assume that for some initial parameters, $\alpha_i = \alpha_i^{(0)}$, an asymmetric steady state $\vecg{u}^{(0)}$
exists. Next, we consider the changed parameters $\alpha_i = \alpha_i^{(0)} + \delta \alpha_i$ \New{and investigate whether this results} in a slightly modified steady state $\vecg{u}= \vecg{u}^{(0)} + \vecg{\delta u}$. We insert both expressions into Eq.~\eqref{eq:stst_general} with $\partial_t u_i=0$ and linearize the result:
\begin{align}
\sum_j \mathcal{D}_{xx}^{(i)}\left(\mathcal{L}_{ij} ~ \delta u_j\right) =& -\delta\alpha_i \mathcal{D}_{xx}^{(i)} \frac{\delta \mathcal{G}_C}{\delta u_i}.
\label{eq:linearized_steadystate}
\end{align}
Now, if we can always find a steady state correction $\vecg{\delta u}$ that solves Eq.~\eqref{eq:linearized_steadystate} then there exists a branch of asymmetric steady states. The solvability condition (which is sufficient) for this to be the case is
\begin{equation}
0\overset{!}{=}  \sum_i \delta \alpha_i  \braket{\delta u_i^\dagger,  \mathcal{D}_{xx}^{(i)} \frac{\delta \mathcal{G}_C}{\delta u_i}} 
\label{eq:solvability}
\end{equation}
with the adjoint zero eigenvector $\vecg{\delta u}^\dagger$ solving the adjoint linear homogeneous equation
\begin{equation}
\sum_j\left(\mathcal{L}^\dagger\right)_{ij} {D}_{xx}^{(j)} \,\delta u_j^\dagger = 0\,\quad \forall i.
\label{eq:linear_dagger}
\end{equation}
The expression $\braket{\dots,\dots}$ denotes \New{a scalar product in function space determined via} spatial integration over the finite domain $\left[-\frac{L}{2},\frac{L}{2}\right]$ with periodic boundary conditions.\footnote{Our analysis also applies to an infinite domain. Domains with other boundary conditions do not allow stationary drifting states and are not considered here.}

\paragraph*{Step 2: Finding the adjoint zero eigenvector}~\\
We can write the adjoint linear operator $\underline{\mathcal{L}}^\dagger$ as
\begin{equation}\label{eq:L_Ldagger}
\left(\mathcal{L}^\dagger\right)_{ij} = \mathcal{L}_{ji}^*= \mathcal{L}^\mathcal{G}_{ii} \delta_{ij} + \alpha_j \mathcal{L}^{\mathcal{G}_C}_{ij}  = \frac{\alpha_j}{\alpha_i} \left(\mathcal{L}^\mathcal{G}_{ii} \delta_{ij} + \alpha_i \mathcal{L}^{\mathcal{G}_C}_{ij}\right) = \frac{\alpha_j}{\alpha_i} \mathcal{L}_{ij} \,,
\end{equation}
where we have used the self-adjointness of $\underline{\mathcal{L}}^\mathcal{G}$ and $\underline{\mathcal{L}}^{\mathcal{G}_C}$. Inserting Eq.~\eqref{eq:L_Ldagger} into Eq.~\eqref{eq:linear_dagger} we obtain
 \begin{align}
\sum_j \frac{\alpha_j}{\alpha_i} \mathcal{L}_{ij} {D}_{xx}^{(j)} \,\delta u_j^\dagger = 0~\nonumber \\
\Leftrightarrow \sum_j  \mathcal{L}_{ij} \alpha_j {D}_{xx}^{(j)} \,\delta u_j^\dagger = 0.
\label{eq:Jdagger}
\end{align}
Comparing the adjoint linear equation \eqref{eq:Jdagger} with Eq.~\eqref{eq:Ldxu0}  we see that a permitted adjoint zero eigenvector can be found as
\begin{equation}\label{eq:adjointEV}
\alpha_j \mathcal{D}_{xx}^{(j)} \delta u_j^\dagger = N \partial_x u^{(0)}_j
\end{equation}
with one arbitrary (normalization) constant $N$. In the variational case $\alpha_j =\alpha$ the adjoint zero eigenvector is the translation mode for the nonconserved fields and the second spatial integral of the translation mode for the conserved fields, respectively. In the nonvariational case each component of the adjoint zero eigenvector is still proportional to the (independent) translation mode of the corresponding field. However, the ratio of these modes is given by the inverse ratio of the (active) coupling parameters, i.e.
\begin{equation}\label{eq:ratio}
\frac{\mathcal{D}_{xx}^{(i)} \delta u_i^\dagger}{\mathcal{D}_{xx}^{(j)} \delta u_j^\dagger} = \frac{\alpha_j}{\alpha_i}\frac{\partial_x u^{(0)}_i}{\partial_x u^{(0)}_j}\,.
\end{equation}
Moreover, the adjoint zero eigenvector has to satisfy the boundary conditions, i.e., $\delta u_i^\dagger(x=\frac{L}{2}) \overset{!}{=} \delta u_i^\dagger(x=-\frac{L}{2})$. This is trivially true for the nonconserved part since $\delta n_i^\dagger = -\frac{N}{\alpha_i}\partial_x n_i^{(0)}$. For the conserved part we integrate Eq.~\eqref{eq:adjointEV} twice and find
\begin{align}
\delta c_i^\dagger(x) = & \frac{N}{\alpha_i} \left( \int^x_{-\frac{L}{2}}{\rm d} x' c_i^{(0)}(x') + E_i x + F_i\right). \label{eq:c_dagger}
\end{align}
Then $\delta c_i^\dagger(x=\frac{L}{2}) \overset{!}{=} \delta c_i^\dagger(x=-\frac{L}{2})$ yields
\begin{align}
\frac{N}{\alpha_i}  \left( \int^\frac{L}{2}_{-\frac{L}{2}}{\rm d} x' c_i^{(0)}(x') + E_i \frac{L}{2} + F_i \right) 
\overset{!}{=}&
\frac{N}{\alpha_i}  \left( \int^{-\frac{L}{2}}_{-\frac{L}{2}}{\rm d} x' c_i^{(0)}(x') - E_i \frac{L}{2} + F_i \right) ~\\
\Rightarrow E_i =& -\frac{1}{L} \int^\frac{L}{2}_{-\frac{L}{2}}{\rm d} x' c_i^{(0)}(x')\,.
\end{align}
Thus the periodic boundary conditions determine the integration constants $E_i$ as the (negative) mean value of the steady conserved field $c_i^{(0)}$. The remaining integration constants $F_i$ remain undetermined.

Note that if some coupling parameters are zero, the factor $\frac{1}{\alpha_i}$ in Eq.~\eqref{eq:L_Ldagger} is problematic. This partially coupled case needs to be considered separately as done in Appendix~\ref{app:partially}.

\paragraph*{Step 3: Showing that the adjoint zero eigenvector solves the solvability condition}~\\
We are left to show that the adjoint zero eigenvector solves the solvability condition~\eqref{eq:solvability}. For this purpose we note that for any functional $\mathcal{G}[\vecg u]=\int {\rm d}x\,f(\vecg u, \partial_x \vecg u, \partial_{xx} \vecg u,\dots)$ 
\begin{equation}
  0=\int {\rm d}x\,\partial_x f =  \braket{\partial_x \vecg{u}, \frac{\delta \mathcal{G}}{\delta \vecg u}}\equiv\sum_i \braket{\partial_x u_i, \frac{\delta \mathcal{G}}{\delta u_i}}.
\label{eq:phixF}
\end{equation}
For the self-interaction terms contained in $\mathcal{G}$ ($=\sum_i \mathcal{G}_i [u_i] $), we therefore have
\begin{equation}
0= \braket{\partial_x u_i, \frac{\delta \mathcal{G}_i}{\delta u_i}}~\, \, \forall~ i\,,
\label{eq:phixF_i}
\end{equation}
here evaluated on the steady state $\vecg{u}=\vecg{u}^{(0)}$.

Next, we take the steady state of Eq.~\eqref{eq:stst_general} and multiply its $i$-th component by the $i$-th component of the adjoint zero eigenvector $\delta u_i^\dagger$, yielding
\begin{align}\label{eq:step3}
0 = & \braket{\delta u^\dagger_i,   \mathcal{D}_{xx}^{(i)} \left( \frac{\delta \mathcal{G}}{\delta u_i} + \alpha_i \frac{\delta \mathcal{G}_C}{\delta u_i }\right)}= \braket{ \mathcal{D}_{xx}^{(i)} \delta u^\dagger_i, \frac{\delta \mathcal{G}}{\delta u_i}} + \alpha_i \braket{ \delta u^\dagger_i, \mathcal{D}_{xx}^{(i)}  \frac{\delta \mathcal{G}_C}{\delta u_i }}\,.
\end{align}
To obtain the first term we twice integrated by parts assuming that the boundary terms vanish.\footnote{For the active PFC model this assumption is not valid. We consider this special case separately in Sec.~\ref{sec:apfc}.}
In the fully coupled case ($\alpha_i \neq 0$ $\forall i$) we insert Eq.~\eqref{eq:adjointEV} for the first term in Eq.~\eqref{eq:step3} and use Eq.~\eqref{eq:phixF_i} to obtain
\begin{equation}\label{eq:coupled}
0=\braket{ \delta u^\dagger_i, \mathcal{D}_{xx}^{(i)}  \frac{\delta \mathcal{G}_C}{\delta u_i }} \quad\forall i\,.
\end{equation}
In view of Eqs.~\eqref{eq:phixF_i}--\eqref{eq:coupled} each summand in Eq.~\eqref{eq:solvability} vanishes, i.e., the adjoint zero eigenvector solves the solvability condition. This result extends to the partially coupled case as well (Appendix \ref{app:partially}).

In summary, we have shown that any asymmetric steady state of \New{the spurious gradient dynamics}~\eqref{eq:stst_general} that exists for a particular set of parameters remains at rest when these parameters are continuously changed (at least over some finite range). In particular, the asymmetric states of the passive system in general do not begin to drift when the system becomes active unless the \New{spurious gradient structure}~\eqref{eq:stst_general} is broken. Evidently this is a necessary but not sufficient requirement for the appearance of motion. Note that the drifting asymmetric states in the models discussed in Sec.~\ref{sec:allmodels} are found on branches that have no counterpart in the passive limit.

\subsection{Onset of motion}\label{sec:motion}

Besides the knowledge that resting states exist, it is also of great interest to determine the onset of motion. For this we follow Ref.~\cite{OpGT2018pre} and consider an expansion in the drift velocity $v$ of the form
\begin{equation}\label{eq:expansion}
\vecg{u}(x,t) = \vecg{u^{(0)}}(x+vt) + v \vecg{u^{(1)}}(x+vt) + \mathcal{O}(v^2)\,,
\end{equation}
as appropriate for steadily drifting states in the vicinity of the drift bifurcation. In the comoving frame $\xi=x+vt$, $\partial_x = \partial_\xi $ and $\partial_t = v \partial_\xi$, and Eq.~\eqref{eq:stst_general} yields
\begin{equation}\label{eq:onset_linear}
\sum_j \mathcal{D}^{(i)}_{xx} \mathcal{L}_{ij} u_j^{(1)}  =  \partial_x u_i^{(0)}
\end{equation}
at leading order in $v$. Here $\partial_x \vecg{u^{(0)}}$ is the Goldstone mode.
This inhomogeneous equation can be solved for $\vecg{u^{(1)}}$ provided the solvability condition
\begin{equation}\label{eq:onset_solv}
0\overset{!}{=}\braket{\partial_x \vecg{u^{(0)}}, \vecg{\delta u}^\dagger} = -\braket{ \vecg{u^{(0)}}, \partial_x \vecg{\delta u}^\dagger} 
\end{equation}
holds, where $\vecg{\delta u}^\dagger$ is the adjoint zero eigenvector determined in the previous section [see Eq.~\eqref{eq:adjointEV}].

For the fully coupled case we have determined in the previous section that
\begin{equation}
\mathcal{D}_{xx}^{(i)} \delta u_i^\dagger = \frac{N}{\alpha_i} \partial_x u_i^{(0)}\,,
\end{equation}
i.e., the nonconserved components are
\begin{equation}
\delta n_i^\dagger =-\frac{N}{\alpha_i} \partial_x n_i^{(0)}
\end{equation}
and the conserved components are
\begin{equation}
\partial_x \delta c_i^\dagger =\frac{N}{\alpha_i}\left( c_i^{(0)} + E_i \right) =   \frac{N}{\alpha_i}\left( c_i^{(0)} -\frac{1}{L} \int_{-\frac{L}{2}}^{\frac{L}{2}}{\rm d} x'\, c_i^{(0)}  \right)\,.
\end{equation}
Then the solvability condition \eqref{eq:onset_solv} becomes
\begin{align}
0=&\sum_{i=1}^{M_1} -\frac{N}{\alpha_i} \braket{\partial_x n_i^{(0)},\partial_x n_i^{(0)}} -
\sum_{i=M_1+ 1}^{M_1+M_2} \frac{N}{\alpha_i} \braket{c_i^{(0)}, c_i^{(0)} + E_i } ~\\
\Leftrightarrow 0 =&\sum_{i=1}^{M_1} \frac{1}{\alpha_i}  \frac{1}{L}\int_{-\frac{L}{2}}^\frac{L}{2} {\rm d} x\, |\partial_x n_i^{(0)}|^2 + \sum_{i=M_1+ 1}^{M_1+M_2}\frac{1}{\alpha_i} \left( \frac{1}{L} \int_{-\frac{L}{2}}^\frac{L}{2} {\rm d} x\, |c_i^{(0)}|^2 - \left(\frac{1}{L} \int_{-\frac{L}{2}}^{\frac{L}{2}} {\rm d} x\, c_i^{(0)}\right)^2 \right)
~\label{eq:solv_1} \\
\Leftrightarrow 0= &\sum_{i= 1}^{M_1} \frac{1}{\alpha_i}\underset{\text{variance}}{\underbrace{ \left(\left< \left(\partial_x n_i^{(0)}\right)^2\right> - \underset{=0}{\underbrace{\left< \partial_x n_i^{(0)}\right>}}^2 \right)}} +  \sum_{i=M_1+ 1}^{M_1+M_2} \frac{1}{\alpha_i} \underset{\text{variance}}{\underbrace{\left(\left< \left(c_i^{(0)} \right)^2  \right> -\left<c_i^{(0)} \right>^2\right)}} \,,
\label{eq:onset_condition}
\end{align}
\New{where in the transition from \eqref{eq:solv_1} to \eqref{eq:onset_condition} we have implicitly defined the mean value via $\left< \dots \right>$.}
Equation~\eqref{eq:onset_condition} tells us that motion sets in at the particular point where the sum of the variances weighted by the corresponding inverse coupling parameter vanishes. These are the variances of the first spatial derivative of the nonconserved fields and those of the conserved fields themselves.

The above prediction relies on the \New{spurious gradient structure}~\eqref{eq:stst_general}, and will be employed in the following section. Note that in the variational case ($\alpha_i=\alpha$) Eq.~\eqref{eq:onset_condition} can never be satisfied since all variances are positive, i.e., there is no onset of motion, an observation in line with expected behavior. The onset of motion in the partially coupled case is discussed in Appendix~\ref{app:partially}.

The expression for the onset of motion calculated above holds for any steady state, i.e., it applies equally to \New{parity-symmetric} and asymmetric steady states; it does not apply to time-dependent states, including steadily drifting states for which condition \eqref{eq:onset_condition} may be fulfilled even though the velocity is nonzero. \New{For steady states with parity symmetry} the bifurcation is a drift-pitchfork bifurcation provided $\vecg u^{(1)}$ breaks the parity symmetry of $\vecg u^{(0)}$. \New{For asymmetric steady states} $\vecg u^{(0)}$ is already asymmetric, and the onset of motion then corresponds to a drift-transcritical bifurcation. Both types of bifurcation can be found in the models considered in Sec.~\ref{sec:allmodels}. Based on the new insights of the present section we now revisit these models.

\section{The models of Section~\ref{sec:allmodels} as spurious gradient dynamics}\label{sec:rewriteModels}

\New{In Sec.~\ref{sec:allmodels} we have considered a number of specific models that all show nongeneric behavior. In the previous section, Sec.~\ref{sec:deriv}, we identified a specific model structure -- the spurious gradient dynamics~\eqref{eq:stst_general} -- that allows for nongeneric steady asymmetric states even in nonvariational systems. Here, we revisit the models of Sec.~\ref{sec:allmodels} and show that they all have the spurious gradient dynamics form.}

\subsubsection{Active phase field crystal model}\label{sec:apfc}
\Resub{At first sight,} the active phase field crystal model~\eqref{eq:dtapfc} does not seem to fit the \New{spurious gradient dynamics form}~\eqref{eq:stst_general}. Since the density-like order parameter $\phi$ is conserved 
its dynamics can be written according to Eq.~\eqref{eq:stst_general} as
\begin{equation}\label{eq:apfc_phi}
 \partial_t \phi = \partial_{xx} \left( \left[\epsilon+\left(1+\partial_{xx}\right)^{2}\right]\phi +\phi^{3}- v_{0} \int^x {\rm d} x'\, P \right)\,.
\end{equation}
The coupling term must have a similar structure in the equation for the polarization field, i.e., we write somewhat unusually
\begin{equation}\label{eq:apfc_pol}
\partial_t P = \partial_{xx} \left( c_1 P + c_2 P^3 - D_{\mathrm{r}}\int^x \int^{x'} {\rm d} x'' {\rm d} x' \left( c_1 P + c_2 P^3 \right)- v_{0} \int^x {\rm d} x'\, \phi \right)\,.
\end{equation}
Now, for the existence of steady asymmetric states it is crucial that $c_2 =0$. Then we can identify 
\begin{align}
\mathcal{G} = &\mathcal{F}^{\text{SH}} + \mathcal{F}^{P} + \frac{1}{2}D_{\mathrm r}c_1 \int_{-\frac{L}{2}}^{\frac{L}{2}} {\rm d}x \left( ~\int^x~{\rm d} x'\, P \right)^2\,,  ~\\
\mathcal{G}_C=&  \int_{-\frac{L}{2}}^{\frac{L}{2}} {\rm d}x \left( P ~\int^x~{\rm d} x' ~\phi\right)\,,~\\
\alpha_1 =& v_0 \quad \text{and} \quad \alpha_2=-v_0\,
\end{align}
and in this way the active PFC model \New{takes on the spurious gradient dynamics form} \Resub{defined by} Eq.~\eqref{eq:stst_general}. 
Furthermore, for steady states $(\phi^{(0)}, P^{(0)})$ we find, via integration of Eq.~\eqref{eq:apfc_pol} over the whole domain, that
\begin{equation}
\int_{-\frac{L}{2}}^{\frac{L}{2}} ~{\rm d} x'\, P^{(0)} = 0\,.
\end{equation}
\New{In analogy to Eq.~\eqref{eq:Ldxu0} we also apply an indefinite spatial integration of Eqs.~\eqref{eq:apfc_phi} and \eqref{eq:apfc_pol} and obtain (still for $c_2=0$),}
\begin{align}
\begin{split}\label{eq:phi_lin} 
0=&\partial_{x} \left( \left[\epsilon+\left(1+\partial_{xx}\right)^{2}\right]\phi^{(0)} +\left(\phi^{(0)}\right)^{3}- v_{0} \int^x {\rm d} x' P^{(0)} \right) ~\\
 & \equiv \mathcal{L}_{11}^\mathcal G \partial_x \phi^{(0)} + v_0 \mathcal{L}_{12}^{\mathcal G_C} \partial_x P^{(0)}\,,  \end{split}~\\
\begin{split}\label{eq:pol_lin}
J_P = &\partial_{x} \left( c_1 P^{(0)} - D_{\mathrm{r}} \int^x\int^{x'} {\rm d} x'' {\rm d} x' c_1 P^{(0)} - v_{0} \int^x {\rm d} x' \phi^{(0)} \right)~\\ \equiv & \mathcal{L}_{22}^\mathcal G \partial_x P^{(0)} - v_0 \mathcal{L}_{12}^{\mathcal G_C} \partial_x \phi^{(0)}\,, 
\end{split}
\end{align}
\New{Here $J_P$ is a constant of integration that is nonzero due to the nonlocal terms in $\mathcal{L}_{22}^\mathcal G $ and $\mathcal{L}_{12}^{\mathcal G_C}$, in contrast to Eq.~\eqref{eq:Ldxu0}. Note that $\mathcal{L}_{21}^{\mathcal G_C}$ also contains a nonlocal term, although the corresponding flux is given by $-v_{0} \int_{-L/2}^{L/2}  {\rm d} x P^{(0)}$ and thus vanishes. The nonzero constant flux $J_P$ is calculated by integrating Eq.~\eqref{eq:pol_lin} over the domain:}
\begin{equation}
J_P=-\frac{D_\mathrm r c_1}{L } \int_{-\frac{L}{2}}^\frac{L}{2} {\rm d} x \int_{-\frac{L}{2}}^x {\rm d} x'\, P^{(0)}(x')- \frac{v_0}{L} \int_{-\frac{L}{2}}^\frac{L}{2}  {\rm d} x \phi^{(0)} \,.
\end{equation}
\New{The nonvanishing flux $J_P$ indicates a difference that is not captured by the calculation in the previous section [cf.~Eq.~\eqref{eq:Ldxu0}]. Despite this the general result remains valid}. First, regarding the adjoint zero eigenvector from Eq.~\eqref{eq:c_dagger} we have 
\begin{align}
\delta\phi^\dagger = \frac{1}{v_0} \left( \int^x_{-\frac{L}{2}} {\rm d} x'\,\phi^{(0)} + E_1 x + F_1\right)\label{eq:phid} \,,~\\
\delta P^\dagger =  - \frac{1}{v_0} \left(\int^x_{-\frac{L}{2}}{\rm d} x'\, P^{(0)} + E_2 x + F_2\right)\label{eq:pold}
\end{align}
with
\begin{align}
E_1=& -\frac{1}{L}\int_{-\frac{L}{2}}^\frac{L}{2} {\rm d} x'\,\phi^{(0)} = -\bar \phi\,,~\\
E_2 =& -\frac{1}{L}\int_{-\frac{L}{2}}^\frac{L}{2} {\rm d} x'\,P^{(0)} = 0\,.
\end{align}
\New{As in the general derivation the constant $F_1$ remains undetermined since the adjoint linear equation only contains first and second derivatives of $\delta\phi^\dagger $. However, owing to the double integral in Eq.~\eqref{eq:apfc_pol} $\delta P^\dagger$ occurs by itself which then determines $F_2$. In particular, the adjoint zero eigenvector solves the adjoint linear problem
\begin{align}
0=& \left( \left[\epsilon+\left(1+\partial_{xx}\right)^{2}\right] +3\left(\phi^{(0)}\right)^{2}\right) \partial_{xx} \delta\phi^\dagger + v_{0}  \partial_x \delta P^\dagger\,,~\label{eq:phiad} \\
0=& c_1 \partial_{xx}\delta P^\dagger   -  D_{\mathrm{r}} c_1 \delta P^\dagger  + v_0 \partial_x \delta \phi^\dagger ~\label{eq:polad}\,.
\end{align}
Inserting expressions \eqref{eq:phid}-\eqref{eq:pold} into the adjoint linear equations \eqref{eq:phiad}-\eqref{eq:polad} and comparing with Eqs.~\eqref{eq:phi_lin}-\eqref{eq:pol_lin} we see that Eq.~\eqref{eq:phiad} is fulfilled. From Eq.~\eqref{eq:polad} we determine
\begin{equation}
F_2= \frac{J_P + v_0 \bar \phi}{D_\mathrm r c_1}= -\frac{1}{L } \int_{-\frac{L}{2}}^\frac{L}{2} {\rm d} x \int_{-\frac{L}{2}}^x {\rm d} x'\, P^{(0)}(x')\,.
\end{equation}
} Second, for step 3 of the derivation in the previous section we need the identity
\begin{equation}\label{eq:problem2_apfc}
0 = \braket{\delta u^\dagger_i,   \mathcal{D}_{xx}^{(i)} \frac{\delta \mathcal{G}}{\delta u_i}}= \braket{ \mathcal{D}_{xx}^{(i)} \delta u^\dagger_i, \frac{\delta \mathcal{G}}{\delta u_i}}\,,
\end{equation}
that is based on vanishing boundary terms that implicitly occur via partial integration. To verify this assumption we note that
\begin{equation}
\partial_{xx} \frac{\delta \mathcal{G}}{\delta P} \sim \partial_{xx}  \int^x \int^{x'} {\rm d} x'' {\rm d} x'' P^{(0)} = P^{(0)}
\end{equation}
so that 
\begin{align}
\braket{\delta P^\dagger,   P^{(0)} }= &\braket{ \int^x_{-\frac{L}{2}}{\rm d} x\, P^{(0)}  + F_2, P^{(0)}} = \int_{-\frac{L}{2}}^\frac{L}{2} {\rm d} x\, \left(\int^x_{-\frac{L}{2}} {\rm d} x'\,P^{(0)} + F_2\right) P^{(0)} ~\nonumber ~\\= &
\frac12\left(\int^\frac{L}{2}_{-\frac{L}{2}} {\rm d} x\, P^{(0)}\right)^2  + F_2 \int^\frac{L}{2}_{-\frac{L}{2}}{\rm d} x\, P^{(0)}= 0
\end{align}
and hence Eq.~\eqref{eq:problem2_apfc} still holds. Thus after these additional considerations the active PFC model with $c_2=0$ also falls into the class of models for which we may expect steady asymmetric states, as already found in Sec.~\ref{sec:aPFCModel}.

In contrast, when $c_2 \neq 0$ we cannot write the kinetic equations in the form of Eq.~\eqref{eq:stst_general} since a term proportional to $\iint {\rm d}x'' {\rm d}x'\, P^3$ cannot be written as a functional derivative. We conclude that for $c_2\neq 0$ all asymmetric states necessarily drift.

To emphasize that resting behavior is not caused by the linearity of Eq.~\eqref{eq:apfc_pol} when $c_2=0$ we also consider the problem
\begin{equation}
\partial_t P = \partial_{xx} \left( c_1 P + c_2 P^3 - D_{\mathrm{r}}c_1 \int^x \int^{x'} {\rm d} x'' {\rm d} x' P- v_{0} \int^x {\rm d} x' \phi \right) \label{eq:dtP2}
\end{equation}
and still observe asymmetric steady states.

According to Eq.~\eqref{eq:onset_condition}, the steady states in either case lose stability to drift when
\New{
\begin{equation}
\left< \left(\phi^{(0)}\right)^2  \right> -\left<\phi^{(0)}\right>^2 - \left< \left(P^{(0)}\right)^2  \right> = -\bar \phi^2+\int_{-\frac{L}{2}}^{\frac{L}{2}} {\rm d} x \,  \left( \left|\phi^{(0)}\right|^2 - \left|P^{(0)}\right|^2\right)  = 0\,. \label{eq:onset_aPFC}
\end{equation}}
This equation predicts correctly the location of the drift-pitchfork and the drift-transcritical bifurcations, indicated by open and filled triangles in Fig.~\ref{fig:aPFC}(a). \New{The specific form \eqref{eq:onset_aPFC} for the case of vanishing mean density was originally derived in Ref.~\cite{OpGT2018pre}.}

\subsubsection{Nonreciprocal Cahn-Hilliard model}
The nonreciprocal Cahn-Hilliard model \eqref{eq:CHCH} can also be written in \New{the spurious gradient dynamics form}~\eqref{eq:stst_general} with
	\begin{align}
	\mathcal{G}=& \mathcal{F}_1^{CH} +  \mathcal{F}_2^{CH} \,, \\
	\text{where  }\qquad \mathcal{F}_i^{CH} =& \int_{-\frac{L}{2}}^{\frac{L}{2}} {\rm d}x  \frac{\kappa_i}{2 } \left(\partial_x \phi_i \right)^2 + f_i (\phi_i) ~\label{eq:F_CH}\,,\\
	\mathcal{G}_C=& -\int_{-\frac{L}{2}}^{\frac{L}{2}} {\rm d}x \phi_1 \phi_2\,, ~\\
	\alpha_1 = & \rho + \alpha, \quad \alpha_2=  \rho - \alpha
	\end{align}
and $\kappa_1 = 1$, $\kappa_2=\kappa$. According to Eq.~\eqref{eq:onset_condition} the onset of motion occurs when a steady state satisfies\footnote{The corresponding expression in \cite{FrTh2021ima} is missing the term $\braket{\phi^{(0)}_i}^2$. }
\begin{align}
\sum_i \frac{1}{\alpha_i} \left(\left< \left(\phi_i^{(0)} \right)^2  \right> -\left<\phi_i^{(0)} \right>^2\right)= &\, 0~\nonumber \\
\Leftrightarrow
\int_{-\frac{L}{2}}^\frac{L}{2} {\rm d}x\,|\phi_1^{(0)}|^2 -\frac{1}{L} \left(\int_{-\frac{L}{2}}^{\frac{L}{2}}{\rm d}x \, \phi_1 ^{(0)}\right)^2 + \frac{\rho + \alpha}{\rho - \alpha} \left( \int_{-\frac{L}{2}}^\frac{L}{2} {\rm d}x \,|\phi_2^{(0)}|^2 -\frac{1}{L} \left(\int_{-\frac{L}{2}}^{\frac{L}{2}} {\rm d}x \,\phi_2 ^{(0)}\right)^2 \right) = &\, 0\,.
\end{align}
This condition correctly predicts the location of the drift-pitchfork bifurcations shown as open triangles in Fig.~\ref{fig:CH+CH_subcritical}(b,c) and as well of the drift-transcritical bifurcation shown as a filled triangle in Fig.~\ref{fig:CH+CH_subcritical}(c). It also confirms that no drift bifurcations occur on the branches shown in Fig.~\ref{fig:CH+CH_snaking}(a,d).

A different nonreciprocal Cahn-Hilliard model is studied in Ref.~\cite{SATB2014c}. After rescaling, this model takes the form
\begin{alignat}{1}
\begin{aligned}
\qquad \frac{\partial\phi_1}{\partial t}  &= \frac{\partial^2}{\partial x^2} \left(- \frac{\partial^2\phi_1 }{\partial x^2} +f_1'(\phi_1)  \right) +\alpha_1 \phi_2\,,\\
\qquad \frac{\partial\phi_2}{\partial t}  &=  \frac{\partial^2}{\partial x^2} \left(-\kappa\frac{\partial^2 \phi_2}{\partial x^2} +f_2'(\phi_2)\right) + \alpha_2 \phi_1 \,.
\end{aligned}
\label{eq:CHCH_baer}
\end{alignat}
Here the nonvariational coupling breaks both conservation laws. Despite this the model can still be written \New{in the spurious gradient dynamics form} \eqref{eq:stst_general} using the functional $\mathcal{G}$ given above and adapting only $\mathcal{G}_C$:
\begin{equation}
\mathcal{G}_C= \int_{-\frac{L}{2}}^{\frac{L}{2}} {\rm d}x \left(\phi_1 \int^x {\rm d}x' \int^{x'} {\rm d}x''  \phi_2 \right)\,.
\end{equation}
Similar to the active PFC model discussed in Sec.~\ref{sec:apfc} integration of the steady state version of Eqs.~\eqref{eq:CHCH_baer} yields
\begin{align}
0= \int_{-\frac{L}{2}}^\frac{L}{2} {\rm d}x\, \phi_1^{(0)} = \int_{-\frac{L}{2}}^\frac{L}{2}  {\rm d}x\,\phi_2^{(0)}\,,
\end{align}
i.e., the mean densities of both fields vanish. 
Further, two nonzero constant fluxes $J_i$ with
\begin{align}
J_1= \frac{\alpha_1}{L} \int_{-\frac{L}{2}}^\frac{L}{2} {\rm d} x \int_{-\frac{L}{2}}^x {\rm d} x' \phi_2^{(0)}(x')\,, ~\\
J_2= \frac{\alpha_2}{L} \int_{-\frac{L}{2}}^\frac{L}{2} {\rm d} x \int_{-\frac{L}{2}}^x {\rm d} x' \phi_1^{(0)}(x') \,
\end{align}
are found.
Again, the nonzero fluxes determine the constants $F_i$ for the adjoint zero eigenvectors [see Eq.~\eqref{eq:c_dagger}]:
\begin{align}
\delta \phi_i^\dagger = \int^x  {\rm d}x'\,\phi_i^{(0)} + F_i = \int^x  {\rm d}x'\, \phi_i^{(0)} + \frac{J_i}{\alpha_i}\,.
\end{align}
\New{In contrast to the active PFC model, here the integral terms do not occur in $\mathcal{G}$, but only in $\mathcal{G}_C$. Thus Eq.~\eqref{eq:problem2_apfc} is trivially fulfilled.} We conclude that our general result also holds for this nonreciprocal nonconserved Cahn-Hilliard model. According to Eq.~\eqref{eq:onset_condition} the onset of motion occurs when
\begin{equation}
\int_{-\frac{L}{2}}^\frac{L}{2}  {\rm d}x\,\left(|\phi_1^{(0)}|^2 + \frac{\rho + \alpha}{\rho - \alpha} |\phi_2^{(0)}|^2\right)=0 \,,
\end{equation}
a prediction that can in principle be checked following the computations in \cite{SATB2014c}.

\subsubsection{Coupled Cahn-Hilliard and Swift-Hohenberg equations}
 
The model~\eqref{eq:CHSH} fits the family of \New{spurious gradient dynamics given by Eq.}~\eqref{eq:stst_general} with
\begin{align}
\mathcal{G}= &\mathcal{F}^{\text{CH}} + \mathcal{F}^{\text{SH}}\,,~\\	
\text{where  }\qquad \mathcal{F}^{\text{SH}} =& \int_{-\frac{L}{2}}^{\frac{L}{2}} {\rm d}x  \left(-\frac{\phi_2}{2} \left[r-\left(1+\partial_{xx}\right)^{2}\right]\phi_2 + f_2(\phi_2)\right)\,, ~\\
	\mathcal{G}_C=& \int_{-\frac{L}{2}}^{\frac{L}{2}} {\rm d}x\, \phi_1 \phi_2\,, ~\\
	\alpha_1 = & \rho + \alpha, \quad \alpha_2=  \rho - \alpha\,
\end{align}
with $\mathcal{F}^{\text{CH}}$ as in Eq.~\eqref{eq:F_CH}. In contrast to the previous models in which all variables exhibit conserved dynamics, here $\phi_1$ is conserved but $\phi_2$ is not. Thus $\phi_1$ takes the role of $c_1$ and $\phi_2$ takes the role of $n_1$ in Eq.~\eqref{eq:stst_general0}.
Since Eq.~\eqref{eq:onset_condition} still applies we see that steady states become unstable with respect to drift when
\begin{equation}
\frac{\rho + \alpha}{\rho - \alpha}  \int_{-\frac{L}{2}}^\frac{L}{2} {\rm d} x\, |\partial_x \phi_2^{(0)}|^2 +  \int_{-\frac{L}{2}}^\frac{L}{2} {\rm d} x\, |\phi_1^{(0)}|^2 - \frac{1}{L} \left( \int_{-\frac{L}{2}}^{\frac{L}{2}} {\rm d} x\, \phi_1^{(0)}\right)^2 = 0\,.
\end{equation}
This condition correctly predicts all drift-pitchfork and drift-transcritical bifurcations shown in Fig.~\ref{fig:CHSH_bifdiag}(a).

\subsubsection{\Resub{Two-species active PFC model}}
If $c_2=0$ we can rewrite Eqs.~\eqref{eq:apfc2} in a similar way to what we have done for the \Resub{one-species} active PFC model in Sec.~\ref{sec:apfc}:
\begin{align}
\partial_{t}\phi_1 &= \partial_{xx}\left[\left[\epsilon+\left(1+\partial_{xx}\right)^{2}\right]\phi_1+\left(\bar{\phi}_1+\phi_1\right)^{3}  + c \phi_2  -v_{0} \int^x {\rm d} x' P \right]\,, ~\\
\partial_{t}P &= \partial_{xx}\left[ c_1 P  - D_{\mathrm{r}}c_1 \int^x \int^{x'} {\rm d} x'' {\rm d} x' P- v_{0} \int^x {\rm d} x' \phi_1 \right]\,, ~\\
\partial_{t}\phi_2 &= \partial_{xx}\left[\left[\epsilon+\left(1+\partial_{xx}\right)^{2}\right]\phi_2+\left(\bar{\phi}_2+\phi_2\right)^{3} + c \phi_1\right] \,.
\end{align}
We see that these equations fit the \New{spurious gradient dynamics} form~\eqref{eq:stst_general} with
\begin{align}
\mathcal{G}=& \sum_{i=1,2} \mathcal{F}_i^{SH} + \mathcal{F}^{P}\,,~\label{eq:PFC_F} \\
\mathcal{F}^{P}=&\int_{-\frac{L}{2}}^{\frac{L}{2}} {\rm d}x \left[ \frac{c_1}{2}P^2  -  \frac{c_1}{2} D_{\mathrm{r}} \left(\int^x {\rm d}x' P\right) ^2 \right]\,, ~\\
\mathcal{G}_C=& \int_{-\frac{L}{2}}^{\frac{L}{2}} {\rm d}x \left[c  \phi_1 \phi_2 +v_0 \left( P ~\int^x~{\rm d} x' ~\phi_1\right)\right]\,, ~\\
\alpha_1 = & \alpha_3 = 1 \quad \text{and} \quad \alpha_2 =- 1  \,.
\end{align}
Similar considerations to those used in the \Resub{one-species} active PFC model regarding the double integral term in the polarization equation apply here as well, and help us understand the existence of asymmetric steady states in this model. Owing to the vanishing of the mean densities and the polarization in steady state, i.e., 
\begin{equation}
\int_{-\frac{L}{2}}^\frac{L}{2} {\rm d}x \, \phi^{(0)}_i = \int_{-\frac{L}{2}}^\frac{L}{2} {\rm d}x \, P^{(0)} = 0\,,
\end{equation}
the onset of motion occurs when the steady state satisfies
\begin{equation}\label{eq:aPFC2_onset}
\int_{-\frac{L}{2}}^\frac{L}{2} {\rm d}x\, \left(|\phi_1^{(0)}|^2 - |P^{(0)}|^2 + |\phi_2^{(0)}|^2 \right) = 0\,,
\end{equation}
cf.~Eq.~\eqref{eq:onset_condition}. This condition confirms that none of the steady states shown in Figs.~\ref{fig:apfc2_snaking_phi0j} and \ref{fig:apfc2_snaking_shiftphi02_0.15_phi0j} exhibit a drift instability. For larger activities such instabilities do occur and are also reliably predicted by Eq.~\eqref{eq:aPFC2_onset} (not shown).

\subsubsection{Reaction-diffusion system}
Remarkably, even the FitzHugh--Nagumo model, Eqs.~\eqref{eq:rd} with $\varepsilon=0$, can be written in the \New{spurious gradient dynamics} form \eqref{eq:stst_general}, with
\begin{align}
\mathcal{G}=& \mathcal{G}_1 + \mathcal{G}_2 = \int_{-\frac{L}{2}}^{\frac{L}{2}}{\rm d}x \left(\frac{D}{2}\left( \partial_x u_1 \right)^2 - \frac{\lambda}{2} u_1^2 + \frac14 u_1^4 - \kappa u_1 + \frac{1}{2} \left( \partial_x u_2 \right)^2 + \frac{1}{2} u_2^2 \right)\,,~\\
\mathcal{G}_C=& \int_{-\frac{L}{2}}^{\frac{L}{2}} {\rm d}x ~u_1 u_2\,,~\\
\alpha_1=& 1 \quad \text{and} \quad \alpha_2=-1 \,.
\end{align}
This observation explains the presence of resting asymmetric states in this model, such as state (d) in Fig.~\ref{fig:rd_snaking}. According to Eq.~\eqref{eq:onset_condition} stationary drift sets in when a steady state satisfies
\begin{equation}
\int_{-\frac{L}{2}}^\frac{L}{2} {\rm d}x\, \left( \left|\partial_x u_1^{(0)}\right|^2 -  \left|\partial_x u_2^{(0)}\right|^2 \right) = 0\,.
\end{equation}
Since none of the steady states shown in Fig.~\ref{fig:rd_snaking} meet this condition, no onset of motion is observed. However, when $\varepsilon \neq 0$ the equation structure~\eqref{eq:stst_general} is broken and consequently all asymmetric states now drift as verified in Fig.~\ref{fig:rd_snaking}(e).

\New{We point out, finally, that there exists a partial overlap between the spurious gradient dynamics structure introduced here and the skew-gradient form discussed in Refs.~\cite{Yana2002jde,KuYa2003pd}. In particular, the FitzHugh-Nagumo model is an example for both structures. However, it appears that none of the other examples introduced in Section~\ref{sec:allmodels} can be brought into the skew-gradient form of Ref.~\cite{KuYa2003pd} and, vice versa, that one can construct examples with the skew-gradient form of Ref.~\cite{KuYa2003pd} that do not correspond to our spurious gradient dynamics structure~\eqref{eq:stst_general}. In contrast, the more restricted skew-gradient structure considered in Ref.~\cite{Yana2002jde} does correspond to a strict subset of the more general form of Eq.~\eqref{eq:stst_general}. 
Nevertheless, our results regarding the existence of stationary asymmetric states (Sec.~\ref{sec:deriv}) also apply to all models with the skew-gradient form of \cite{KuYa2003pd} as their equation for steady states can be brought into the form for steady states of the spurious gradient dynamics. } 

\section{How is generic behavior restored? }\label{sec:restored}

Now that we have been able to explain the presence of asymmetric steady states in models with certain nongeneric couplings, let us show how generic behavior may be recovered. For this purpose, we consider a model consisting of two nonlinearly coupled Swift-Hohenberg (SH) equations. First, we verify that steady asymmetric states are still observed for nonlinear couplings that respect the specific structure \New{of spurious gradient dynamics as given by Eq.~\eqref{eq:stst_general}} and considered in the previous section. Subsequently, we show that all steady asymmetric states of this type immediately begin to move when this structure is broken. We then use the insight from the study of this model to suggest different ways in which each of the models of Sec.~\ref{sec:allmodels} can be amended to restore generic behavior.

We begin by introducing two (nonlinearly) coupled Swift-Hohenberg equations as a simple example that captures the properties of resting states in active systems that either arise from symmetry or from the special form of \New{spurious gradient dynamics in Eq.~\eqref{eq:stst_general}}:
\begin{alignat}{1}
\begin{aligned}
\partial_t \phi_1 = & \left[r_1-\left(q_1^2+\partial_{xx}\right)^{2}\right]\phi_1 + \delta_1 \phi_1^2 + b_1 \phi_1^{3} -\phi_1^5+ \alpha_1\left( \phi_2 + \beta \phi_1 \phi_2^2 + \gamma \partial_{xxx} \phi_2 \right)\,,  \\
\partial_t \phi_2 = & \left[r_2 -\left(q_2^2+\partial_{xx}\right)^{2}\right]\phi_2 + \delta_2 \phi_2^2 +b_2 \phi_2^{3} -\phi_2^5 + \alpha_2\left( \phi_1 + \left(\beta + \widetilde{\beta}\right)  \phi_1^2 \phi_2 - \left(\gamma + \widetilde \gamma\right) \partial_{xxx} \phi_1 \right)\, .
\end{aligned}
\label{eq:coupledSH}
\end{alignat}
Based on our general considerations we expect to find steady asymmetric states whenever $\widetilde \beta=\widetilde \gamma=0$, since Eqs.~\eqref{eq:coupledSH} then take the form of two nonconserved fields that can be written in the form~\eqref{eq:stst_general} with
\begin{align}
\mathcal{G}=& \mathcal{G}^{\text{SH}}_1 + \mathcal{G}^{\text{SH}}_2\,,~\\
\mathcal{G}^{\text{SH}}_i =& \int_{-\frac{L}{2}}^{\frac{L}{2}} {\rm d}x  \left(-\frac{\phi_i}{2} \left[r_i-\left(q_i^2+\partial_{xx}\right)^{2}\right]\phi_i + \frac{\delta_i}{3} \phi_i^3 - \frac{b_i}{4} \phi_i^{4} + \frac16 \phi_i^6\right)\,,~\\
\mathcal{G}_C=& - \int_{-\frac{L}{2}}^{\frac{L}{2}} {\rm d}x \left(\phi_1 \phi_2 + \frac{\beta}{2}\phi_1^2 \phi_2^2 + \gamma \phi_1  \partial_{xxx} \phi_2 \right)\,.
\end{align}

\begin{figure}[h!]
	\includegraphics[width=\textwidth]{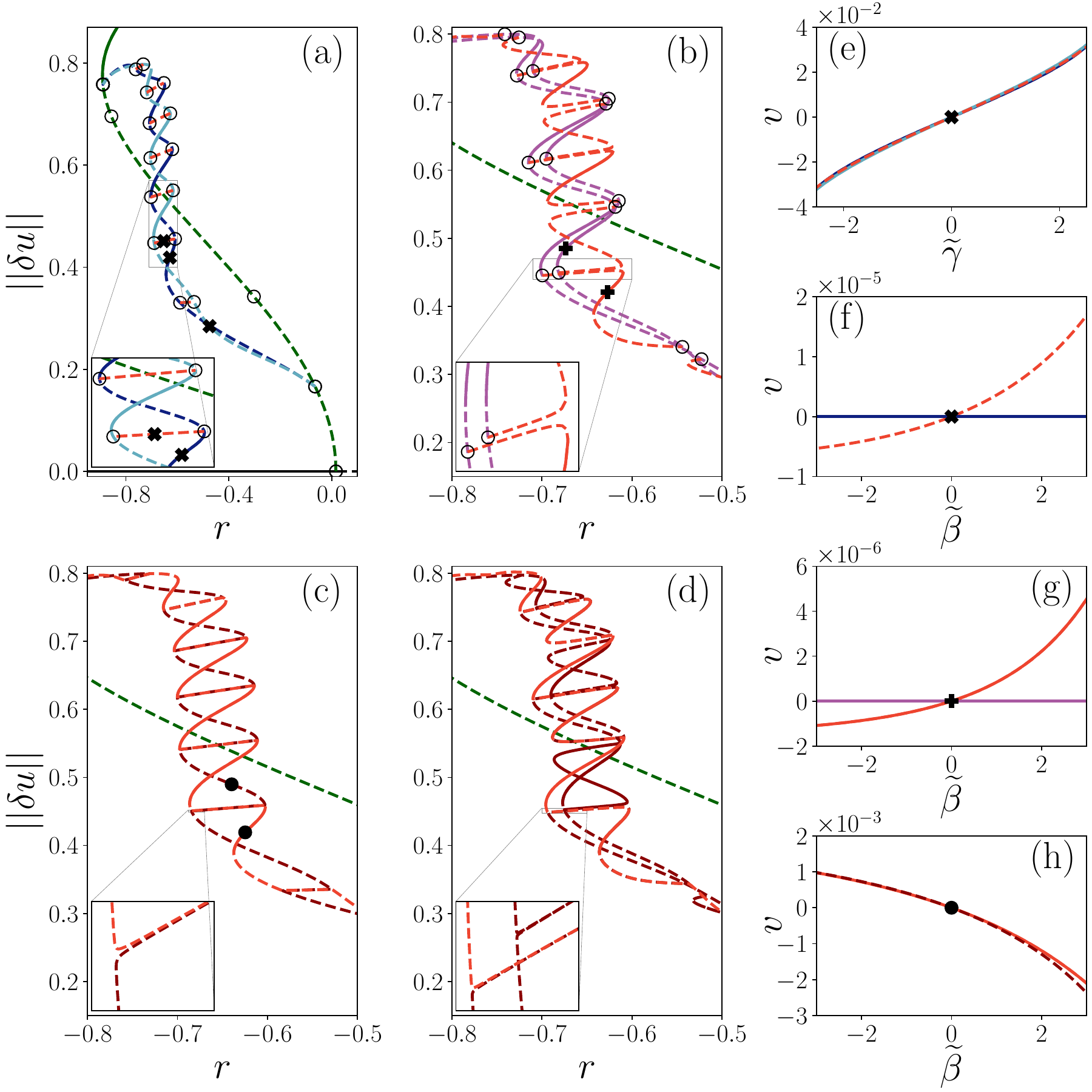}
	\caption{\small Two coupled SH equations \eqref{eq:coupledSH} exhibit snaking behavior with resting symmetric and asymmetric states when $\widetilde \beta = \widetilde \gamma=0$. Shown is the L$_2$ norm \eqref{eq:norm} in the case of (a) reflection and inversion symmetry ($\gamma=\delta_i=0$), (b) reflection symmetry but broken inversion symmetry ($\delta_1=0.02$, $\delta_2=0.04$, $\gamma=0$), (c) inversion symmetry but broken reflection symmetry ($\delta_i=0$, $\gamma=0.8$) and (d) broken reflection and broken inversion symmetry ($\delta_1=0.02$, $\delta_2=0.04$, $\gamma=0.8$), all as a function of the parameter $r$. Panels (e)-(h) show the velocity $v$ of specific localized states marked by cross, plus and circle symbols in panels (a), (b) and (c), respectively, as a function of $\widetilde \gamma$ (e) and $\widetilde \beta$ [(f), (g) and (h)]. Red and dark red lines indicate steady asymmetric states. The remaining parameters are $r_\Delta=-0.3$, $q_1=1$, $q_2=1.3$, $\alpha_1=-0.1$, $\alpha_2=0.125$, $\beta=0.2$. The domain size is $L=10 \pi$. Line styles and symbols are as in Table~\ref{tab:instab}. }
	\label{fig:cSH_snaking} 
\end{figure}

If, in addition, we turn off the nonlinear and dispersive coupling terms $(\beta=\gamma=0)$, we obtain two {\it linearly} coupled Swift-Hohenberg equations similar to the SH equations studied in Ref.~\cite{SATB2014c} where the authors focus on the onset of motion but do not report the existence of any steady asymmetric states. Beside the additional coupling terms we also consider the possibility of breaking the inversion symmetry, i.e., taking $\delta_i \neq 0$, but employ a cubic-quintic nonlinearity to allow for a subcritical primary bifurcation even in the inversion-symmetric case. Note that we permit the wavenumbers $q_i$ to differ. The choice of these wavenumbers does not affect the symmetry properties of the model but can accentuate the consequences of changing the parameters that do (see Fig.~\ref{fig:cSH_snaking} below).

We begin with the reflection and inversion-symmetric case, i.e., we set $\delta_1=\delta_2=\gamma=\widetilde \gamma=0$. Furthermore we set $\widetilde \beta=0$. Equations~\eqref{eq:coupledSH} are then symmetric with respect to the reflection $\mathcal{R}: (x,\phi_1,\phi_2) \mapsto (-x,\phi_1,\phi_2)$ and the inversion $\mathcal{I}: (\phi_1,\phi_2) \mapsto (-\phi_1,-\phi_2)$. Note that the gradient structure is broken when $\alpha_1$ and $\alpha_2$ have opposite signs although the form of Eq.~\eqref{eq:stst_general} is maintained as $\widetilde \beta=\widetilde \gamma=0$. Furthermore, we define $r_1=r$ and $r_2=r+ r_\Delta$ and use $r$ as the main control parameter. In many applications of the SH equation $r$ plays the role of an effective temperature.

The bifurcation diagram in Fig.~\ref{fig:cSH_snaking}(a) confirms that steady asymmetric states exist and form the rung states of the snakes-and-ladders structure typical of SH-like equations. At the primary bifurcation of the trivial state (black line) a periodic state with 5 peaks emerges subcritically (dark green line). This state is further destabilized in the first secondary bifurcation generating two distinct branches of localized states that ultimately form the snaking structure and eventually terminate back on the 5-peak branch. Note that each of the two snaking branches corresponds to two overlapping branches of symmetry-related states \cite{HoKn2011pre}: the light blue [dark blue] branch consists of states that are invariant under $\mathcal{R}$ and hence of even parity [point-symmetric states invariant under $\mathcal{R} \circ \mathcal{I}$ and hence of odd parity]. Both snaking branches change their stability at successive folds and nearby pitchfork bifurcations. The latter give rise to branches of asymmetric states (red dashed lines), i.e., states that are neither reflection nor point-symmetric. Note that these states are at rest even though the gradient structure is broken. They cannot, however, be realized dynamically since they are always unstable.

Next, we investigate whether the different localized states start to move when $\widetilde \beta$ or $\widetilde \gamma$ are nonzero. We consider the states at the locations denoted by the three cross symbols in panel~(a) and vary $\widetilde \gamma$ and $\widetilde \beta$ in panels~(e) and (f, g, h), respectively. For $\widetilde \gamma \neq 0$ the additional dispersive coupling term $\widetilde \gamma  \partial_{xxx} \phi_i$ breaks the reflection symmetry {\it and} \New{the spurious gradient dynamics form of} Eq.~\eqref{eq:stst_general}. In this case all states lose their even or odd parity symmetry and start to move [cf.~the three colored lines in panel~(d)] with the velocity increasing monotonically with increasing $\widetilde \gamma$. This is similar to the behavior in the one-field case when reflection symmetry is broken \cite{BuHK2009pre}. Note that in panel (e) the speed of the originally symmetric states only depends on $|\widetilde \gamma|$ while the drift of the originally asymmetric states is (weakly) sensitive to the sign of $\widetilde \gamma$. However, this effect is not visible to the eye.

In panel~(f) we see that for nonzero $\widetilde \beta$ the asymmetric state begins to move (red dashed line), in contrast to the even and odd symmetric localized states (dark blue line) which remain at rest for any $\widetilde \beta$. Since the additional terms $\sim \phi_i \phi_j^2$ preserve both inversion and reflection symmetry, the states with the corresponding symmetries remain at rest even though the form \eqref{eq:stst_general} of the model is broken. The absence of movement due to symmetries is explained in Sec.~\ref{sec:generic}. This behavior is also reported in Ref.~\cite{BuDa2012sjads} where nonvariational terms are added to the original SH equation. There, the model retains reflection symmetry and, hence, all symmetric states are steady for arbitrary parameter values.

Next, we investigate the role of the inversion symmetry $\mathcal{I}$. Panel (b) shows a bifurcation diagram with the same parameters as in (a) except that we turn on the quadratic terms by setting $\delta_1=0.02$ and $\delta_2=0.04$ thereby breaking the inversion symmetry. In particular, the small values of $\delta_1$ and $\delta_2$ allow a direct comparison with the symmetric case in panel~(a). It is observed that the original snaking structure splits at the pitchfork bifurcations of the odd parity states (see inset) resulting in $Z$- and $S$-shaped branches (red lines) as in \cite{HoKn2011pre}. These pitchfork bifurcations correspond in (a) to spontaneous breaking of the point symmetry at the transition from antisymmetric \New{(odd parity symmetry)} to asymmetric \New{(no parity symmetry)} states. In (b) this symmetry is no longer present and so cannot be broken. Instead, all antisymmetric states in (a) become asymmetric in (b). Note that sections of stable asymmetric solutions can be found on the $Z$-shaped structures, which in (a) represent the stable regions of the odd parity states.

The spatial reflection symmetry $\mathcal{R}$ is still preserved in (b), so that even parity states and their pitchfork bifurcations to the asymmetric states persist. However, compared to (a), the degeneracy due to inversion symmetry is now removed and the symmetric states break up into two branches (light purple). These are the RLS$_\text{even}$ and RLS$_\text{odd}$ states discussed earlier (recall that the terms even and odd now refer to the number of peaks and not to the symmetry of the state under $\mathcal{R}$). If we increase the coefficients of the quadratic terms, the folds of the $S$- and $Z$-shaped structures annihilate and one recovers snaking behavior similar to that in (a) although the corresponding states are qualitatively different, just as in the passage from the cubic-quintic to the quadratic-cubic SH equation \cite{HoKn2011pre}. In this context, however, it is striking that we have identified stable resting but asymmetric solutions, a finding we can clearly attribute to the \New{spurious gradient dynamics form} of the problem, Eq.~\eqref{eq:stst_general}. If this form is broken while keeping the reflection symmetry, we see in panel~(g) that the asymmetric states drift, while the symmetric ones remain at rest. Here the speed of the asymmetric states depends strongly on the sign of $\widetilde \beta$.

In panel~(c) we show a bifurcation diagram for $\gamma \neq 0$, but $\delta_i=0$, i.e., this case represents the case of broken reflection symmetry while keeping the inversion symmetry. Here the even and odd parity states from (a) both lose their symmetry and become asymmetric but remain at rest because the form \eqref{eq:stst_general} is preserved. \New{This is another striking aspect of the spurious gradient structure \eqref{eq:stst_general} of the problem which not only prevents motion arising from spontaneous parity symmetry breaking, but also that arising from forced parity symmetry breaking, here represented by $\gamma\ne0$.} As a result of the broken reflection symmetry all pitchfork bifurcations become imperfect and the intertwined branches reconnect (see inset) resulting in a branch of purely unstable asymmetric states (dashed dark red line) and a second branch of alternately stable and unstable asymmetric states (red line). Both branches are degenerate due to the inversion symmetry of the model and so represent states with both signs of asymmetry. If the \New{spurious gradient dynamics} structure \eqref{eq:stst_general} is broken via $\widetilde \beta \neq 0$ all asymmetric states start to drift [see panel~(h)]; the sign of $v$ is selected by the asymmetry of the state considered.

Finally, in panel~(d) we break both inversion and reflection symmetries by setting $\gamma =0.8$, $\delta_1=0.02$ and $\delta_2=0.04$, thereby destroying the degeneracy due to inversion symmetry present in panel~(c) and that due to the reflection symmetry in panel (b). Thus all pitchfork bifurcations of the even parity states in (b) also become imperfect in (d) (see inset). As a result the connections between the different states are changed compared to the other cases. On the one hand there is the red branch that consistently snakes upwards, resembling the red branch in panel~(c). In contrast to panel~(c) where the dark red branch behaves analogously, here the dark red branch first heads upwards but then makes a loop downwards before resuming its upward growth. This is difficult to see in the figure since the splitting of the original odd parity branch is quite small. As a result all localized states (red and dark red lines) are asymmetric but, despite all the broken symmetries, they are still at rest. This is because for vanishing $\widetilde \beta$ and $\widetilde \gamma$ the \New{spurious gradient structure is still preserved}.

The above examples illustrate clearly the considerations that are required to determine whether or not a particular model will exhibit generic behavior. When applied to the model system studied in Sec.~\ref{sec:aPFCModel} these considerations explain why the addition of the term $c_2P^3$ to Eqs.~\eqref{eq:dtapfc} results in drifting asymmetric states [Fig.~\ref{fig:aPFC}(e)]. For the nonreciprocal Cahn-Hilliard model (Sec.~\ref{sec:cCH}) related higher order interaction terms play a similar role, except in special cases, as illustrated by the role of the parameters $\beta$ and $\widetilde \beta$ in Eqs.~\eqref{eq:coupledSH}. \New{Examples where generic behavior is restored are the nonreciprocal Cahn-Hilliard models considered in Refs.~\cite{SaAG2020prx,FrTh2023arxiv}. In Ref.~\cite{SaAG2020prx} the model includes a reciprocal quartic nonlinearity in the free energy and the derived model in Ref.~\cite{FrTh2023arxiv} features all interspecies interaction terms up to cubic order in the order parameter fields. Similarly, one can prevent a spurious gradient dynamics form} for the (linearly) coupled Cahn-Hilliard and Swift-Hohenberg equations in Sec.~\ref{sec:CH+SH}. There one could also imagine a different coupling, a linear coupling that breaks the conservation of the Cahn-Hilliard field. Such a coupling also restores generic behavior. The coupled passive and active PFC model~\eqref{eq:apfc2} \Resub{results from an approximation to a more general PFC model whose structure is systematically derived} from dynamical density functional theory in Ref.~\cite{VHKW2022msmse}. The general model not only includes nonlinear terms in the polarization field (corresponding to $c_2 \neq 0$) but also features nonlinear variational coupling terms between all three fields and variational gradient coupling terms involving the densities. Based on our theoretical study it is clear that any variational coupling term between the polarization field and any of the density fields will restore generic behavior, while any variational coupling between the densities, including either nonlinearities or gradients, would fail to do so. Furthermore, as discussed for the active PFC model without passive particles, the presence of resting asymmetric states (red branches in Figs.~\ref{fig:apfc2_snaking_phi0j} and \ref{fig:apfc2_snaking_shiftphi02_0.15_phi0j}) also requires the absence of nonlinearities in the nonconserved part of the mixed conserved and nonconserved dynamics of the polarization field $P$ (i.e., $c_2=0$).
Finally, we now also understand why the nonlinear coupling $\varepsilon u_2^3$ in the reaction-diffusion model in Eq.~\eqref{eq:rd} precludes resting asymmetric states.

Based on our earlier examples one might now suppose that linear coupling terms always yield nongeneric behavior provided they respect some basic requirements such as the conservation law of the uncoupled field. This is indeed so for two-component models, but not for models with more components. Consider, e.g., a three-component reaction-diffusion model with linear cross-couplings of the form
\begin{align}
\partial_t u : & ~~~\alpha v + \beta w~\\
\partial_t v : & ~~~\alpha' u + \gamma w~ = \frac{\alpha'}{\alpha}\left(\alpha u + \frac{\alpha}{\alpha'} \gamma w \right)\\
\partial_t w : & ~~~\beta' u + \gamma' v = \frac{\beta'}{\beta}\left(\beta u + \frac{\beta}{\beta'} \gamma' v \right)~\,.
\end{align}
For $\beta = \gamma' = 0$ one obtains a three-component FitzHugh-Nagumo model with a second activator field $w$~\cite{YoKK2015pre}. Despite the linearity of the coupling terms the nongeneric \New{spurious gradient structure}~\eqref{eq:stst_general} further demands that $\frac{\alpha}{\alpha'} \gamma = \frac{\beta}{\beta'}\gamma'$, a relation that does not need to hold in general.


\section{Conclusion}\label{sec:conc}

We have seen that a number of commonly used nonvariational models for active matter and reaction-diffusion systems are nongeneric in the sense that they admit steady asymmetric states. \New{Although all the examples we considered evolve in one spatial dimension the nongeneric behavior we have identified can also be found in higher spatial dimensions. Then, besides the resting asymmetric states analogous to the corresponding states in one dimension, we also expect resting states that are asymmetric in two and more spatial directions. For the active PFC model in two dimensions examples of such structures are presented in Figs.~9 and~10 of Ref.~\cite{OKGT2021pre}.}

\New{Evidently, the statement ``asymmetry in the presence of activity necessarily results in a drift'' is not always correct, although we still expect it to hold for one-component models.  For multicomponent models we have identified the spurious gradient dynamics structure as the origin of the unexpected, nongeneric behavior. This structure extends the common gradient dynamics structure responsible for stationary states (including asymmetric ones) in passive models to a related structure that encompasses certain nonvariational and hence active models. Despite its nongenericity the spurious gradient dynamics can be present in multicomponent models with conserved, nonconserved and mixed dynamics, usually as a result of a simple, e.g.,~leading-order, coupling between different order parameter fields. As a consequence, we expect other active, multicomponent models to exhibit nongeneric behavior and hope that the present work will provide guidance for identifying such models. For example the spurious gradient dynamics structure is also present in a subclass of Keller-Segel models as in the one studied in Ref.~\cite{PeGo2015pre}. Particularly noteworthy is our observation that steady asymmetric states can form via both spontaneous parity breaking bifurcations \textit{and} via forced parity breaking as in Sec.~\ref{sec:restored}, i.e., that such states can be present even in models that themselves are not parity-symmetric. In both cases generic active models exhibit drifting asymmetric states only.}

\New{Moreover, we have also seen that in the generic case symmetric standing oscillations and asymmetric traveling states emerge from symmetric steady states in, respectively, Hopf and drift-pitchfork bifurcations. These bifurcations continue to exist in nongeneric models with the spurious gradient structure. However, when asymmetric steady states are present their drift is instead associated with a drift-transcritical bifurcation while the Hopf bifurcation generates oscillating asymmetric states that do not drift. We have employed the spurious gradient structure to derive an expression that predicts the onset of motion via both drift-pitchfork and drift-transcritical bifurcations, and verified its predictions quantitatively in several of the models studied here. Owing to these bifurcations time-periodic states tend to dominate the system behavior at large activity, i.e., at large activity such models behave more and more like generic models.}

\New{We cannot exclude the possibility of further generalizations of the spurious gradient dynamics structure, for example, to nonvariational models with time-delayed feedback. Specifically one could determine if these models can also exhibit nongeneric behavior and, if not, why time-delayed feedback always restores generic behavior.} \New{Moreover, the partial overlap between our spurious gradient dynamics structure and the skew-gradient form of Refs.~\cite{Yana2002jde,KuYa2003pd} merits further study.}

It is evident that in order to avoid models with nongeneric behavior one must pay heed to the considerations detailed in the present work. Several models of active systems in common use suffer from nongeneric behavior and their properties are therefore very sensitive to the inclusion of small additional terms that restore genericity. We believe that this is a serious shortcoming of such models and that models with generic behavior are always preferable, particularly when relating the predictions of such models to experimental observations. It is striking that nongeneric behavior occurs even in the FitzHugh--Nagumo model that is not, in contrast to the other examples, \New{considered as an active matter model. We have seen that the nongeneric character of this model derives from the linear nonreciprocal, activator-inhibitor interaction satisfying the spurious gradient dynamics structure. In fact, purely linear couplings in any two-species reaction diffusion model (without cross diffusion) always result in nongeneric behavior, although this is normally not the case for models with more than two components. Presumably this fact has remained hidden because the affected states are usually unstable, cf.~Fig.~\ref{fig:rd_snaking}.}

\acknowledgments

TFH, MPH, SVG and UT acknowledge support from the doctoral school ‘‘Active living fluids’’ funded by the German-French University (Grant No. CDFA-01-14). EK acknowledges support from the National Science Foundation (Grant No. DMS-1908891). In addition, TFH wishes to thank the foundation “Studienstiftung des deutschen Volkes” for financial support. TFH and MPH contributed equally to this work.


\appendix

\New{\section{General parity symmetry}\label{sec:general_parity}
We consider a general system that is described by an $M$-component order parameter field $\vecg u(\vec x, t)=\left(u_1(\vec x,t), \dots, u_M(\vec x, t)\right)$ with
\begin{align} 
\partial_t \vecg u = \vecg F[\vecg u, \vec \nabla]\,,
\label{eq:active_general}
\end{align}
where $\vec x\in \mathbb{R}^d$ and the $M$-component function $\vecg F$ involves both linear and nonlinear (integro)-differential operators, represented by positive and negative powers of $\vec \nabla$. To consider resting and steadily drifting states, we assume that Eq.~\eqref{eq:active_general} is translation-invariant, i.e., that $\vecg F$ does not explicitly depend on $\vec x$. Hence, we neglect any pinning effects.
As in Eq.~\eqref{eq:steady}, resting and steadily drifting states $\vecg{u^{(0)}}$ solve 
\begin{equation}\label{eq:v=0_a}
0 = \vecg F[\vecg{u^{(0)}}(\vec \xi\,), \vec \nabla_\xi] + \vec v \cdot \vec \nabla_\xi \vecg{u^{(0)}}(\vec \xi\,)\,,
\end{equation}
 where $\vec \nabla_\xi=\left(\partial_{\xi_1},\dots,\partial_{\xi_d}\right)$ indicates derivatives in the comoving coordinates.
Now, each component $u_i(\vec x, t)$ is separated into a parity-symmetric and parity-antisymmetric part (cf.~Eq.~\eqref{eq:uSuA}).
A steady state can have a general (possibly mixed) parity symmetry, i.e., 
$\vecg{u^{(0)}}(-\vec \xi\,) = \mathbf{r} \vecg{u^{(0)}}(\vec \xi\,)\equiv \left(\mathsf{r}_1 u^{(0)}_1(\vec \xi\,),\mathsf{r}_2 u^{(0)}_2(\vec \xi\,), \dots, \mathsf{r}_M u^{(0)}_M(\vec \xi\,)\right) $ with $\mathsf{r}_i \in \left\{ -1, +1 \right\} $, i.e., each component $u^{(0)}_i(\vec \xi\,)$ is either symmetric ($u^{(0)}_i(\vec \xi\,)=S u^{(0)}_{i,\text{S}}(\vec \xi\,)$) or antisymmetric ($u^{(0)}_i(\vec \xi\,)=A u^{(0)}_{i,\text{A}}(\vec \xi\,)$). Each possible combination encoded in the diagonal matrix $\mathbf{r}$ is a representation of the parity symmetry. In particular, if $\mathsf{r}_i= +1$ [$\mathsf{r}_i= -1$]~$\forall i$ we say that ${\vecg{u^{(0)}}}(\vec \xi\,)$ has even [odd] parity symmetry. 
}

\New{
The flow field determines which parity symmetry representations occur for steady states. Mathematically speaking, if $\vecg F[\vecg u, \vec \xi]$ is equivariant under the parity symmetry transformation $\mathcal{R}:(\vec \xi, \vecg u)\to (-\vec \xi,\mathbf{r} \vecg u)$, i.e., if $\vecg F[\mathbf{r} \vecg u, -\vec \nabla_\xi]=\mathbf{r}  \vecg F[\vecg u, \vec \nabla_\xi]$, steady states with parity symmetry $\mathbf{r}$ may exist. 
If $\vecg F$ is equivariant with respect to other symmetry transformations, steady states with various parity symmetry representations may exist. A typical example is that $\vecg F$ is equivariant with respect to reflection $\vec \xi\to-\vec \xi$ and inversion $\vecg u\to -\vecg u$, i.e., $\vecg F[\vecg u, \vec \nabla_\xi]=\vecg F[\vecg u, -\vec \nabla_\xi]=-\vecg F[-\vecg u, \vec \nabla_\xi]$. Then the model exhibits steady states with even and odd parity symmetry. } 

\New{Next, we apply the considerations of Sec.~\ref{sec:generic} to a general parity-symmetric steady state. We apply a spatial reflection to the steady state in Eq.~\eqref{eq:v=0_a}, here for a steady state of any parity symmetry, and obtain
\begin{align}
0=& \vecg F[\vecg{u^{(0)}}(-\vec \xi\,), -\vec \nabla_\xi] - \vec v \cdot \vec \nabla_\xi \vecg{u^{(0)}}(-\vec \xi\,)~ ~\nonumber \\
=& \vecg F[\mathbf{r} \vecg{u^{(0)}}(\vec \xi\,), -\vec \nabla_\xi] -\vec v \cdot \vec \nabla_\xi \mathbf{r} \vecg{u^{(0)}}(\vec \xi\,)~ ~\nonumber \\
=&  \mathbf{r} \left( \vecg F[\vecg{u^{(0)}}(\vec \xi\,), \vec \nabla_\xi] -\vec v \cdot \vec \nabla_\xi \vecg{u^{(0)}}(\vec \xi\,) \right)\nonumber \\
\Rightarrow 0 =&  \vecg F[\vecg{u^{(0)}}(\vec \xi\,), \vec \nabla_\xi] -\vec v \cdot \vec \nabla_\xi \vecg{u^{(0)}}(\vec \xi\,)\,.\label{eq:v=0_b}
\end{align}
A comparison of Eq.~\eqref{eq:v=0_b} with Eq.~\eqref{eq:v=0_a} shows that $\vec v=0$, i.e., steady states with (any) parity symmetry do not drift. Hence, tracking steady states through parameter space we expect to find branches of resting parity-symmetric states since symmetries are unchanged along continuous branches. This relation applies to most of the steady states in this paper, as all primary pattern-forming states\footnote{These are states that emerge from bifurcation of the homogeneous steady state.} are necessarily parity-symmetric and many of the secondary bifurcations also respect this symmetry.}

\New{Next, we consider an asymmetric state that can be decomposed in a parity-symmetric part $\vecg{u^{(0)}}(\vec \xi\,)$ and a small perturbation $\mu \vecg{u^{(1)}}(\vec \xi\,)$, $\mu\ll 1$, with opposite parity symmetry that is responsible for the overall asymmetry, i.e.,
	\begin{equation}
	\vecg{u^{(0)}}(-\vec \xi\,) = \mathbf{r} \vecg{u^{(0)}}(\vec \xi\,)~\,,\qquad \vecg{u^{(1)}}(-\vec \xi\,)= -\mathbf{r}\vecg{u^{(1)}}(\vec \xi\,)\,.
	\end{equation}
Expanding the steady state Eq.~\eqref{eq:v=0_a} in $\mu$, we obtain
\begin{equation}\label{ap:symm1}
0= \vecg F[\vecg{u^{(0)}}(\xi), \vec \nabla_\xi]  + \mu  \tens J[\vecg{u^{(0)}}(\xi), \vec \nabla_\xi]  \vecg{u^{(1)}}(\xi) + \vec v \cdot \vec \nabla_\xi \left(\vecg{u^{(0)}}(\vec \xi) + \mu \vecg{u^{(1)}}(\vec \xi)\right) + \mathcal{O}(\mu^2)\,.
\end{equation}
where $\tens J[\vecg{u^{(0)}}(\vec \xi\,), \vec \nabla_\xi] \equiv \nabla_{\vecg u} \vecg F[\vecg{u^{(0)}}(\vec \xi\,), \vec \nabla_\xi] $ is the $M\times M$-dimensional Jacobi matrix, a differential operator in spatial representation.
Applying the parity transformation $\vec \xi\to -\vec \xi$ gives
\begin{align}
0= & \vecg F[\vecg{u^{(0)}}(-\vec\xi), -\vec \nabla_\xi]  + \mu\,\tens J[\vecg{u^{(0)}}(-\vec \xi), -\vec \nabla_\xi]  \vecg{u^{(1)}}(-\vec \xi) - \vec v \cdot \vec \nabla_\xi \left(\vecg{u^{(0)}}(-\vec \xi) + \mu \vecg{u^{(1)}}(-\vec \xi)\right) + \mathcal{O}(\mu^2\,,)\nonumber~\\
  {\rm or} & \nonumber\\
0= &  \mathbf{r} \vecg F[\vecg{u^{(0)}}(\vec \xi), \vec \nabla_\xi] - \mathbf{r} \mu\,\tens J[\vecg{u^{(0)}}(\vec \xi), \vec \nabla_\xi] \vecg{u^{(1)}}(\vec \xi) - \vec v \cdot \vec \nabla_\xi \left(\mathbf{r} \vecg{u^{(0)}}(\vec \xi) - \mathbf{r} \mu \vecg{u^{(1)}}(\vec \xi)\right) + \mathcal{O}(\mu^2)\,, \label{ap:symm1b}
\end{align}
where in the second step the symmetries of $\vecg{u^{(0)}}$, $\vecg{u^{(1)}}$ and $\vecg F$ are used. Furthermore, $\tens J[\vecg{u^{(0)}}(-\vec \xi\,), -\vec \nabla_\xi]= \nabla_{ \mathbf{r} \vecg u} \vecg F[ \mathbf{r} \vecg{u^{(0)}}(\vec \xi\,), -\vec \nabla_\xi]= \mathbf{r} \mathbf{r}^{-1}\nabla_{\vecg u} \vecg F[\vecg{u^{(0)}}(\vec \xi\,), \vec \nabla_\xi]= \tens J[\vecg{u^{(0)}}(\vec \xi\,), \vec \nabla_\xi]$.
Comparison of Eqs.~\eqref{ap:symm1} and \eqref{ap:symm1b} yields
\begin{equation}\label{ap:symm3}
0= \mu \,\tens J[\vecg{u^{(0)}}(\vec \xi), \vec \nabla_\xi]  \vecg{u^{(1)}}(\vec \xi) + \vec v \cdot \vec \nabla_\xi \vecg{u^{(0)}}(\vec \xi) + \mathcal{O}(\mu^2)\,.
\end{equation}
Multiplying Eq.~\eqref{ap:symm3} from the left with $\vec \nabla_\xi \vecg{u^{(0)}}(\vec \xi\,)$ and integrating over the whole domain gives
\begin{equation}\label{ap:symm5}
 v_i = - \mu \frac{\left< \partial_{\xi_i} \vecg{u^{(0)}} |\tens J \vecg{u^{(1)}}\right>}{\left<\partial_{\xi_i} \vecg{u^{(0)}}| \partial_{\xi_i} \vecg{u^{(0)}}\right>}=-\mu \frac{\left< \tens J^\dagger \partial_{\xi_i} \vecg{u^{(0)}} | \vecg{u^{(1)}}\right>}{\left<\partial_{\xi_i} \vecg{u^{(0)}}| \partial_{\xi_i} \vecg{u^{(0)}}\right>}\,,
\end{equation}
where $\left<\dots\right>$ denotes a scalar product, i.e., a scalar product in order parameter space followed by integration over the domain; $\partial_{\xi_i} $ denotes the spatial derivative in the $\xi_i$ direction.
Consequences of the result in Eq.~\eqref{ap:symm5} are discussed in Sec.~\ref{sec:generic} of the main text.
}

\section{Existence of steady asymmetric states for partially coupled systems}\label{app:partially}
In Sec.~\ref{sec:deriv} we showed that stationary asymmetric states of Eq.~\eqref{eq:stst_general} that exist for a particular parameter set stay at rest for infinitesimally shifted parameters. However, the derivation implicitly assumes that the coupling parameters $\alpha_i$ are nonzero, since otherwise some expressions diverge, see e.g.~Eqs.~\eqref{eq:L_Ldagger} and~\eqref{eq:c_dagger}.  However, this case is essential since, as already explained, Eq.~\eqref{eq:stst_general} represents a passive model as long as all coupling parameters have the same sign. In other words, setting a coupling parameter to zero defines the transition from a passive to an active system. Moreover, since steady asymmetric states are a natural feature of passive systems, an understanding of the transition to an active system is crucial.

We assume, therefore, that one specific coupling parameter is zero, i.e., $\alpha_k=0$ for some $k$ and $\alpha_i\neq 0$ for all $i\neq k$. Note that then the dynamics of the corresponding field $u_k$ is uncoupled, i.e., it is not influenced by the other state variables $u_i$ although the latter do depend on $u_k$. We call the resulting case a partially coupled system. 

The main difference in the partially coupled case relies on to the linearized equations and the construction of the adjoint zero eigenvector. From Eq.~\eqref{eq:Ldxu0} we find that
\begin{equation}
0= \mathcal{L}^\mathcal G_{kk}\partial_x u^{(0)}_k\,.
\end{equation}
The adjoint linear equation \eqref{eq:linear_dagger} is thus solved by
\begin{equation}\label{eq:adjointEV_app}
\mathcal{D}_{xx}^{(j)} \delta u_j^\dagger = N \partial_x u^{(0)}_j \delta_{jk}\,,
\end{equation}  
where $N$ is a normalization constant. In other words, the adjoint zero eigenvector has only one nonzero entry and represents the single translation mode for $u^{(0)}_k$ (or its second integral if the field is conserved). In fact, this result represents the appropriate limit of the adjoint zero eigenvector found in the fully coupled case [Eq.~\eqref{eq:adjointEV}] since for $\alpha_k \to 0$ the ratio $\frac{\mathcal{D}_{xx}^{(k)} \delta u_k^\dagger}{\mathcal{D}_{xx}^{(i)} \delta u_i^\dagger}$ diverges [see Eq.~\eqref{eq:ratio}].

What is not affected by setting $\alpha_k=0$ is that we need to show that the solvability condition~\eqref{eq:solvability} is satisfied for any parameter shift $\delta\alpha_i$, $\forall i$. From the steady state equations for the coupled fields it follows that
\begin{equation}\label{eq:Gc_ui_app}
0=\braket{\partial_x u^{(0)}_i, \frac{\delta \mathcal{G}_i}{\delta u_i}} + \alpha_i \braket{\partial_x u^{(0)}_i, \frac{\delta \mathcal{G}_C}{\delta u_i}} \Rightarrow \braket{\partial_x u^{(0)}_i, \frac{\delta \mathcal{G}_C}{\delta u_i}}= 0 ~\ \forall i\neq k \,.
\end{equation}
This equation, together with the identity~\eqref{eq:phixF}, leads to
\begin{equation}\label{eq:ident_app}
0= \sum_{i\ne k}  \braket{\partial_x u_i^{(0)}, \frac{\delta \mathcal{G}_C}{\delta u_i}} + \braket{\partial_x u_k^{(0)}, \frac{\delta \mathcal{G}_C}{\delta u_k}}= \braket{\partial_x u^{(0)}_k, \frac{\delta \mathcal{G}_C}{\delta u_k}}\,,
\end{equation}
i.e., the solvability condition \eqref{eq:solvability} for the adjoint zero eigenvector \eqref{eq:adjointEV_app}. As a result Eq.~\eqref{eq:solvability} holds even when $\delta \alpha_k \neq 0$, showing that steady states still remain at rest when the system transitions into an active one.

\textit{Onset of motion:}~\\
The onset of motion in the partially coupled case (for one zero coupling parameter) follows from Eq.~\eqref{eq:onset_solv}. Employing the adjoint zero eigenvector given by Eq.~\eqref{eq:adjointEV_app} this condition reduces to
\begin{align}
  0\overset{!}{=} & \begin{cases}
\partial_x n_k^{(0)}\quad \text{if}\, k\leq M_1~\\
\left< \left(c_k^{(0)} \right)^2  \right> -\left<c_k^{(0)} \right>^2 \quad \text{if}\, k> M_1\,.
\end{cases}
 \label{eq:onset_partuncoupled}
\end{align}
Recall again that the partially coupled case represents the transition from the passive to the active system, i.e., if Eq.~\eqref{eq:onset_partuncoupled} is fulfilled when $\alpha_k=0$ then this point represents a higher-codimension bifurcation point at which a whole branch of drifting asymmetric states emerges when $\alpha_k$ passes zero. In \cite{FrWT2021pre} [see Scenario SubMinus] this situation leads to the emergence of two drift-pitchfork and two Hopf bifurcations, and one fold.


%
\end{document}